\documentclass{aa}  
\usepackage{graphicx}
\usepackage{txfonts}
\usepackage[draft]{hyperref}

\usepackage[colorinlistoftodos]{todonotes} 

\usepackage{capt-of}
\usepackage{amsmath}
\usepackage{placeins}
\usepackage[utf8]{inputenc}
\usepackage{filecontents}
\usepackage{natbib}

\newcommand{\hZthinbest}[0]{0.28}
\newcommand{\hZthickbest}[0]{1.12}
\newcommand{\rhodmbest}[0]{0.018}
\newcommand{\erhodmbest}[0]{0.002}
\newcommand{\chisqbest}[0]{2.6}
\newcommand{\Sigmabaryonbest}[0]{44.8}

\begin{document} 

    \title{The vertical force in the Solar Neighbourhood using red clump stars in TGASxRAVE}
    \subtitle{Constraints on the local dark matter density}
    \titlerunning{$K_z$ in the solar neighbourhood}
    \author{Jorrit H.J. Hagen and Amina Helmi} 
    \authorrunning{J.H.J. Hagen \& A. Helmi}
    \institute{Kapteyn Astronomical Institute, University of Groningen,
              Landleven 12, 9747 AD Groningen, The Netherlands\\
              \email{hagen@astro.rug.nl}}
    \date{Received Month XX, 20YY; accepted Month XX, 20YY}

 
  \abstract
   {}
   {We investigate the kinematics of red clump stars in the Solar
     Neighbourhood by combining data from TGAS and RAVE to constrain
     the local dark matter density.}
   {After calibrating the absolute magnitude of red clump stars, we
     characterize their velocity distribution over a radial distance
     range of $6$-$10$ kpc and up to $1.5$ kpc away from the Galactic
     plane. We then apply the axisymmetric Jeans equations on subsets
     representing the thin and thick disks to determine the
     (local) distribution of mass near the disk of our Galaxy.}
   {Our kinematic maps are well-behaved permitting a straightforward
     local determination of the vertical force, which we find to be
     $K_z^{\rm thin} = -2454 \pm 619$ and $K_z^{\rm thick} = -2141 \pm
     774$~$(\mathrm{km/s})^2 \! /\mathrm{kpc}$ at 1.5 kpc away from the
     Galactic plane for the thin and thick disk samples and for thin
     and thick disk scale heights of $\hZthinbest$ kpc and
     $\hZthickbest$ kpc respectively. These measurements can be
     translated into a local dark matter density
     \mbox{$\rho_\textrm{DM} \sim \rhodmbest \pm \erhodmbest$
       $M_\odot/\textrm{pc}^3$}. The systematic error on this estimate
     is much larger than the quoted statistical error, since even a
     10\% difference in the scale height of the thin disk leads to a
     30\% change in the value of $\rho_\textrm{DM}$, and a nearly
     equally good fit to the data.}
   {}


   \maketitle
%



\section{Introduction}
With the launch of the Gaia satellite a wealth of new data is becoming
available on the motions and positions of stars in the Milky Way and
its satellite galaxies \citep{GaiaDR1summary2016,
  GaiaMission2016}. For example, the Tycho Gaia Astrometric Solution
(TGAS) has provided significantly improved proper motions and
parallaxes of nearby stars. The power of these data increases even
further when combined with spectroscopic surveys such as RAVE \citep{Kunderetal2017}, APOGEE \citep{Majewskietal2017}
and LAMOST \citep{Cuietal2012}, as this provides knowledge of the full phase-space
distribution of stars near the Sun, as shown in e.g.
\citet[][]{AllendePrietoetal2016, Helmietal2017, Liangetal2017,
  Monarietal2017, Yu&Liu2018}.

New kinematic maps of the Solar neighbourhood can be used for example,
to obtain more precise estimates of the local dark matter density
$\rho_{\mathrm{DM}}$. Most modern measurements of $\rho_{\mathrm{DM}}$
seem to be consistent with a value just below $\sim 0.01
M_{\odot}/\textrm{pc}^3$ when assuming a total baryonic surface mass
density $\Sigma_{\mathrm{baryon}}$ of $55 M_{\odot}/\textrm{pc}^2$
according to \citet{Read2014}. \citet{McKeeetal2015} argue for a value
of $0.013 M_{\odot}/\textrm{pc}^3$ for $\Sigma_{\mathrm{baryon}} =
47.1 M_{\odot}/\textrm{pc}^2$, and \citet{Bienaymeetal2014} find
$\rho_{DM} = 0.0143 \pm 0.0011 M_{\odot}/\textrm{pc}^3$ for
$\Sigma_{\mathrm{baryon}} = 44.4 \pm 4.1 M_{\odot}/\textrm{pc}^2$
using red clump stars in RAVE DR4. Recently \citet{Sivertssonetal2017}
determined a dark matter density of $0.012 M_{\odot}/\textrm{pc}^3$
using SDSS-SEGUE G-dwarf stars for $\Sigma_{\mathrm{baryon}} = 46.85
M_{\odot}/\textrm{pc}^2$. These estimates are consistent with those
inferred from studies that model the mass of the Milky Way globally
\citep[e.g.][]{Piffletal2014}.

Despite the apparent good agreement between the various values
reported in the literature, local dark matter density estimates are
affected by different ``systematic'' uncertainties \citep[see
e.g.][]{Silverwoodetal2016}. These include of course,
uncertainties on $\Sigma_{\mathrm{baryon}}$, which depend on the gas
mass density which has typically an error of 50\%, and the local
stellar densities although these are typically determined with 10\%
accuracy or better \citep[e.g.][]{Holmberg&Flynn2000,
  Bovy2017}. Furthermore, the presence of multiple populations, and
assumptions made regarding their distribution \citep{MoniBidin2012b,
  Hessman2015, Budenbenderetal2015}, or deviations from equilibrium,
such as bending or breathing of the disk \citep{Widrowetal2012,
  Williamsetal2013} may also affect the conclusions reached
\citep{Baniketal2017}.

In this paper we use TGAS and RAVE data to derive an estimate of the
local dark matter density and explore the impact of uncertainties
in (some of) the characteristic parameters of the Galactic thin and thick disks.  In
Sect. \ref{sec:data} we present the data and selection criteria used
in this work, as well as the new kinematic maps of the Solar
Neighbourhood. In Sect. \ref{sec:dmdensity} we present our local dark
matter measurement and the impact of the disk parameters on the
estimate. We conclude in Sect. \ref{sec:conclusions}.



\section{Data}
\label{sec:data}

\begin{figure*}[ht!] \centering
  \includegraphics[width=0.49\textwidth]{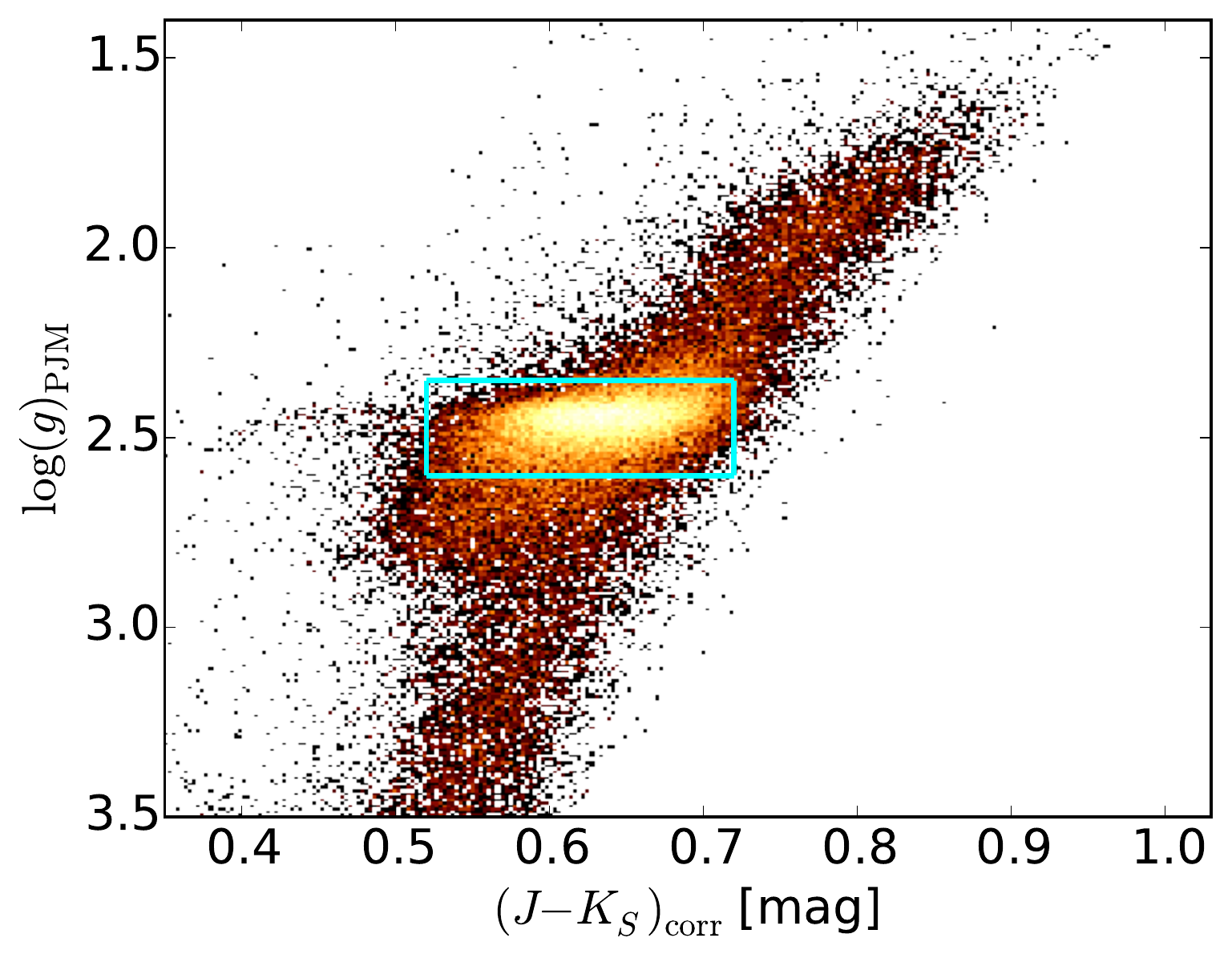}%
  \includegraphics[width=0.49\textwidth]{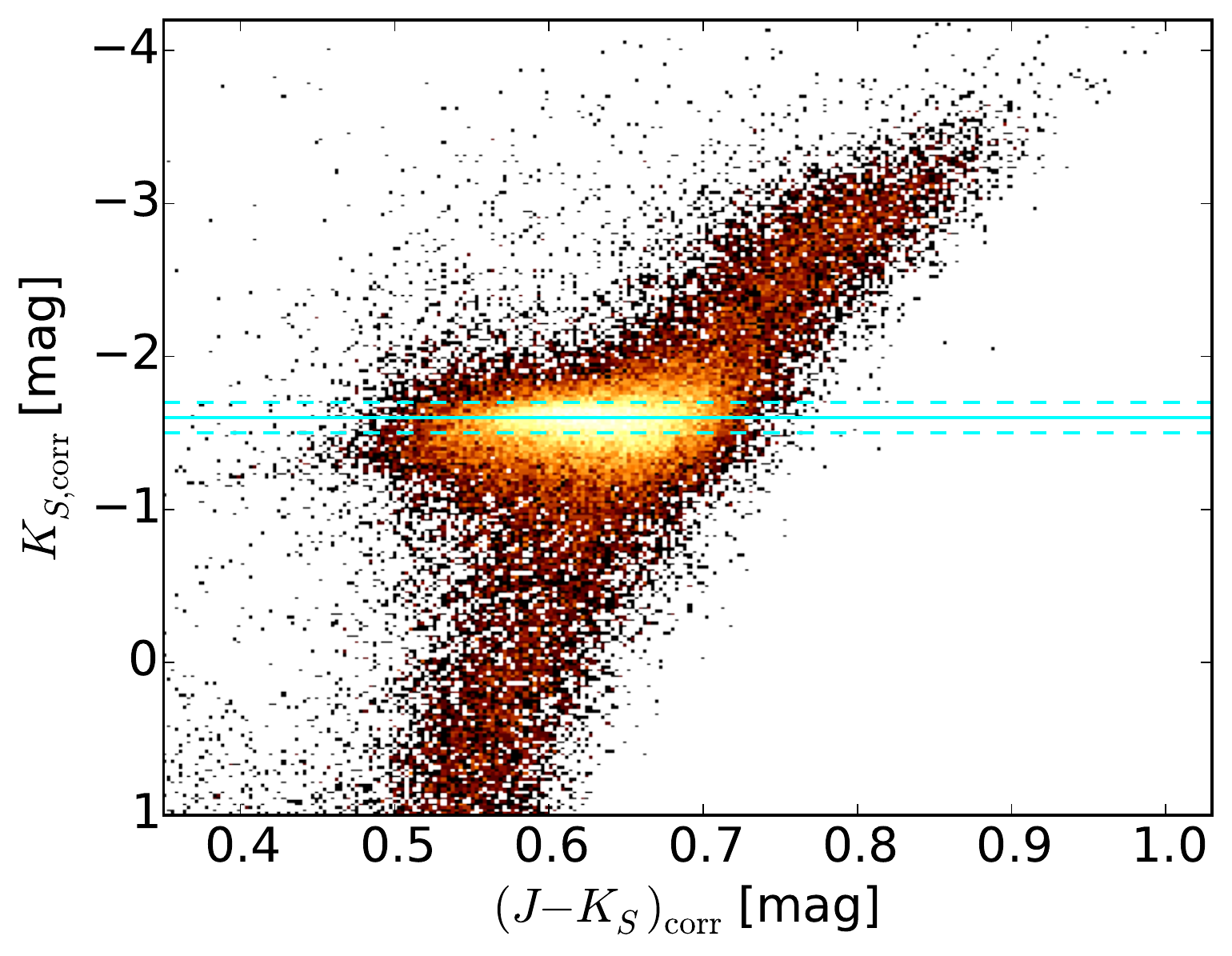}%
  \caption{Selection and calibration of red clump (RC) stars. Left: Distribution of giant stars in our sample in the extinction corrected colour vs. surface gravity space. The blue lines enclose our RC sample. Right: Extinction corrected HR diagram of stars in our sample. The solid and dashed light blue lines indicate respectively, the adopted mean and standard deviation of the red clump absolute $K_S$-band magnitude.} 
  \label{fig:RCselection} 
\end{figure*}%

The dataset we use stems from the cross-match of TGAS to the 5th Data
Release of the Radial Velocity Experiment \citep[RAVE DR5,
][]{Kunderetal2017}. The TGAS dataset contains $\sim 2$ million stars
that are in common between the Gaia DR1 catalogue and the Hipparcos
and Tycho-2 catalogues and provides accurate positions on the sky,
mean proper motions and parallaxes. The RAVE survey is a
magnitude-limited survey ($9<I<12$) of stars in the southern
hemisphere ($\sim 450000$ stars), that for low galactic latitudes ($b
< 25^{\circ}$) uses a colour criterion, $J - K_s \geq 0.5$, which
preferentially selects giant stars.  RAVE provides radial velocities,
astrophysical parameters, as well as a spectro-photometric
parallaxes. The overlap between RAVE DR5 and TGAS contains $\sim
250000$ stars with full phase space information. \citet[PJM2017
hereafter]{McMillanetal2017} used the TGAS astrometric parallax
measurements as priors to derive \textit{improved} spectro-photometric
parallaxes and astrophysical parameters. This is the dataset that we
use here, after applying a few extra quality cuts. We keep only those
stars that have \mbox{\texttt{SNR\_K} $> 20$}, an \texttt{ALGO\_CONV}
parameter equal to either $0$ or $4$, \mbox{\texttt{eHRV} $< 8$ km/s},
and \mbox{\texttt{flag\_any} $= 0$}. This leaves us with a `reliable'
sample containing $108679$ stars. The median distance of stars in this
sample is of $\sim 0.5$ kpc, and their median relative distance error
is $\sim 13$\%.

\subsection{Selecting a red clump stars' sample}
\label{sec:RCsample}
To gain more in distance accuracy we use red clump (RC) stars, as they
act as standard candles \citep[e.g.][]{Paczynski&Stanek1998,
  Groenewegen2008, Girardi2016}. We select RC stars on the basis of
the surface gravity $\log(g)_{\textrm{PJM}}$ and the extinction
corrected 2MASS bands $J_\textrm{corr}$ and $K_{S,\textrm{corr}}$,
where \mbox{$J_\textrm{corr} = J - A_J$} and \mbox{$K_{S,\textrm{corr}} = K_S -
A_{K_S}$}, and $A_J = 0.282 \, A_V$ and $A_{K_S} = 0.112 \, A_V$, where
$A_V$ is taken from PJM2017. Finally we define our `RC sample' by:
\begin{equation}
   0.52 \leq (J - K_S)_\textrm{corr} \leq 0.72 \;\;\; {\rm and} \;\;\;
   2.35 \leq \log(g)_{\textrm{PJM}} \leq 2.60,
\end{equation}
and which contains $26653$ stars.
We calibrate the RC absolute $K_s$-band magnitude by considering a
subsample of $3211$ RC stars with a maximum relative parallax error of
$10\%$. We find these stars to have a mean absolute $K_s$-band
magnitude of $M^{\textrm{RC}}_{K_S} = -1.604$ mag and a
dispersion of $0.064$ mag, which we round-off to
$M^{\textrm{RC}}_{K_S} = -1.60$ mag and a dispersion of $0.1$ mag
(which translates into a $\sim 5$\% error in distance) in the
remainder of this work. These values depend slightly on the maximum
parallax error imposed for the calibration, and they are consistent
with \citet{Hawkinsetal2017}, who find $M^{\textrm{RC}}_{K_S} = -1.61 \pm 0.01$ mag
and a dispersion of $0.17 \pm 0.02$, and \citet{Ruiz-Dernetal2017} who
find $M^{\textrm{RC}}_{K_S} = -1.606 \pm 0.009$ mag, both works mainly using TGAS parallaxes 
and APOGEE spectroscopy. 

In the left panel of Fig. \ref{fig:RCselection} we show the
distribution of giant branch stars in our sample in the extinction
corrected $(J-K_s) _\textrm{corr}$ colour versus surface gravity
$\log(g)_{\textrm{PJM}}$ space. Our selection of red clump stars is
given by the light blue box. In this region there will be some
contamination of red giant branch (RGB) stars since RAVE does not
provide asteroseismic information that can be used to discriminate
between RC and RGB stars \citep[e.g.][]{Beddingetal2011,
  Chaplin&Miglio2013}. According to \citet{Girardi2016} the typical
contamination fraction is of order $30\%$ times the $\log(g)$ boxwidth
in dex used to define the RC sample, thus resulting in a contamination
estimate of $7.5\%$ for our RC selection criteria. In the right-hand
side panel of Fig. \ref{fig:RCselection} we show a similar region of
an extinction corrected HR-diagram by using the PJM2017 improved parallaxes 
for the stars from the reliable sample. The mean and standard deviation 
of the adopted RC absolute $K_S$-band magnitude are plotted as the horizontal 
solid and dashed lines respectively.

\subsection{General properties of the RC sample}
\label{sec:kinematicmaps} 

\begin{figure} \centering 
\includegraphics[width=8.5cm]{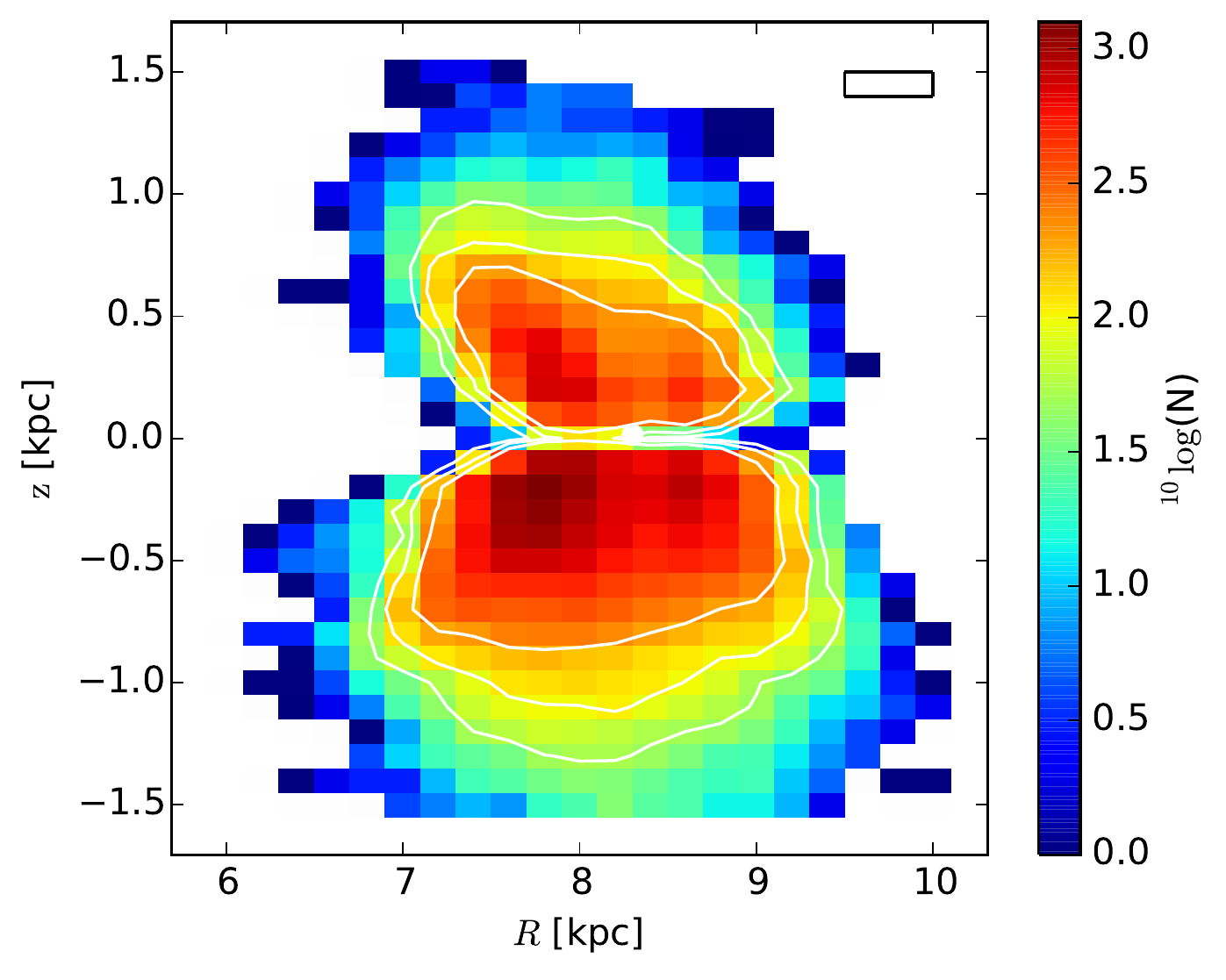} 
\caption{Meridional plane RC star counts in bins of $0.5$~kpc in $R$
  and $0.1$~kpc in $z$ (as indicated by the box in the upper right
  corner). The white contours indicate where the number of RC stars in
  the bins have dropped to $200$, $100$, and $50$ from inner to outer
  contours respectively. The white dot marks the position of the
  Sun. This figure is based on the default realization in which the
  absolute magnitudes of all RC stars are set to
  $M^{\textrm{RC}}_{K_S} = -1.60$~mag.}
 \label{fig:Nstars_2d} 
\end{figure}

Now that we have calibrated the mean absolute magnitude of RC stars
and its spread, we proceed to compute distances to our stars. We here
define a `default realization' in which all RC stars are assumed to
have $M^{\textrm{RC}}_{K_S} = -1.60$~mag.  The resulting spatial
distribution of stars is shown in Fig. \ref{fig:Nstars_2d}, computed
in bins with widths $0.5$~kpc in $R$ and $0.1$~kpc in $z$, after
setting $R_\odot = 8.3$~\text{kpc} \citep{Schonrich2012} and $z_\odot
= 0.014$~\text{kpc} \citep{Binneyetal1997}, for the position of the
Sun with respect to the Galactic centre. This figure shows that our
sample covers $R \sim 6 - 10$~kpc and $|z|\lesssim 1.5$~kpc.  Since
the RAVE survey is more oriented to the southern and inner part of the
Galaxy, there are more RC observed in these regions. 




\begin{figure*}
  \centering
  \includegraphics[width=0.365\textwidth]{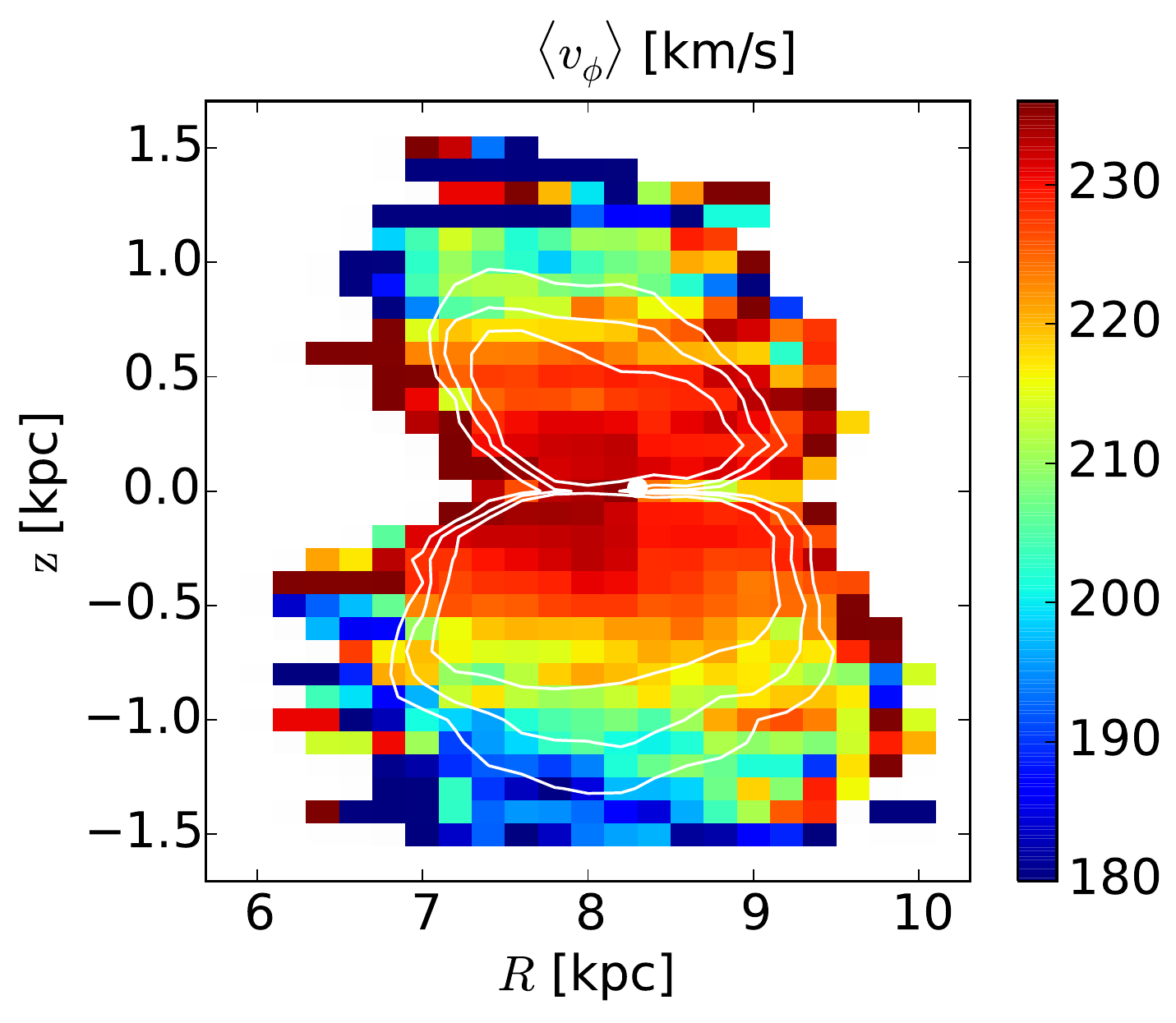}%
  \includegraphics[width=0.31\textwidth]{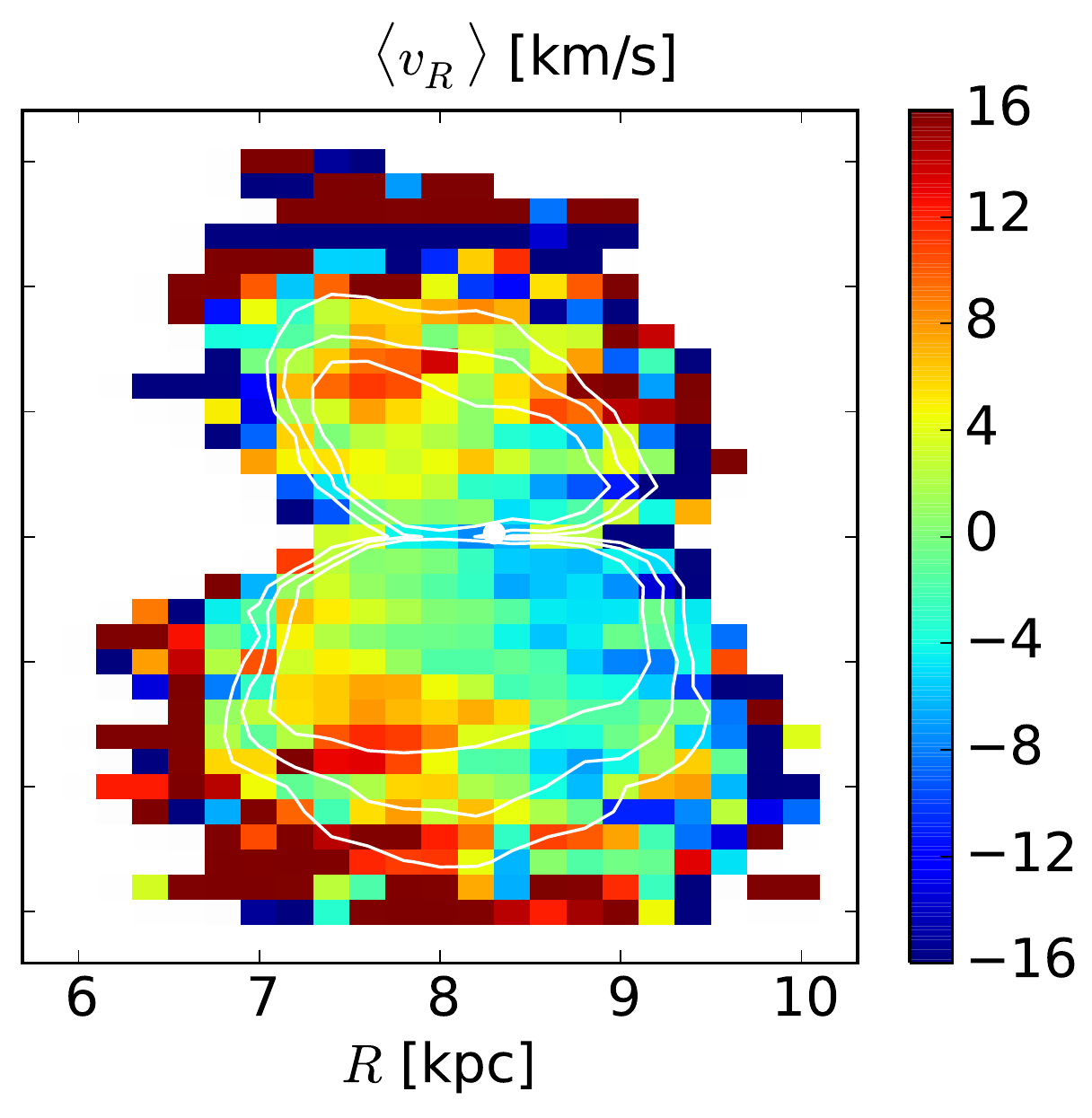}%
  \includegraphics[width=0.31\textwidth]{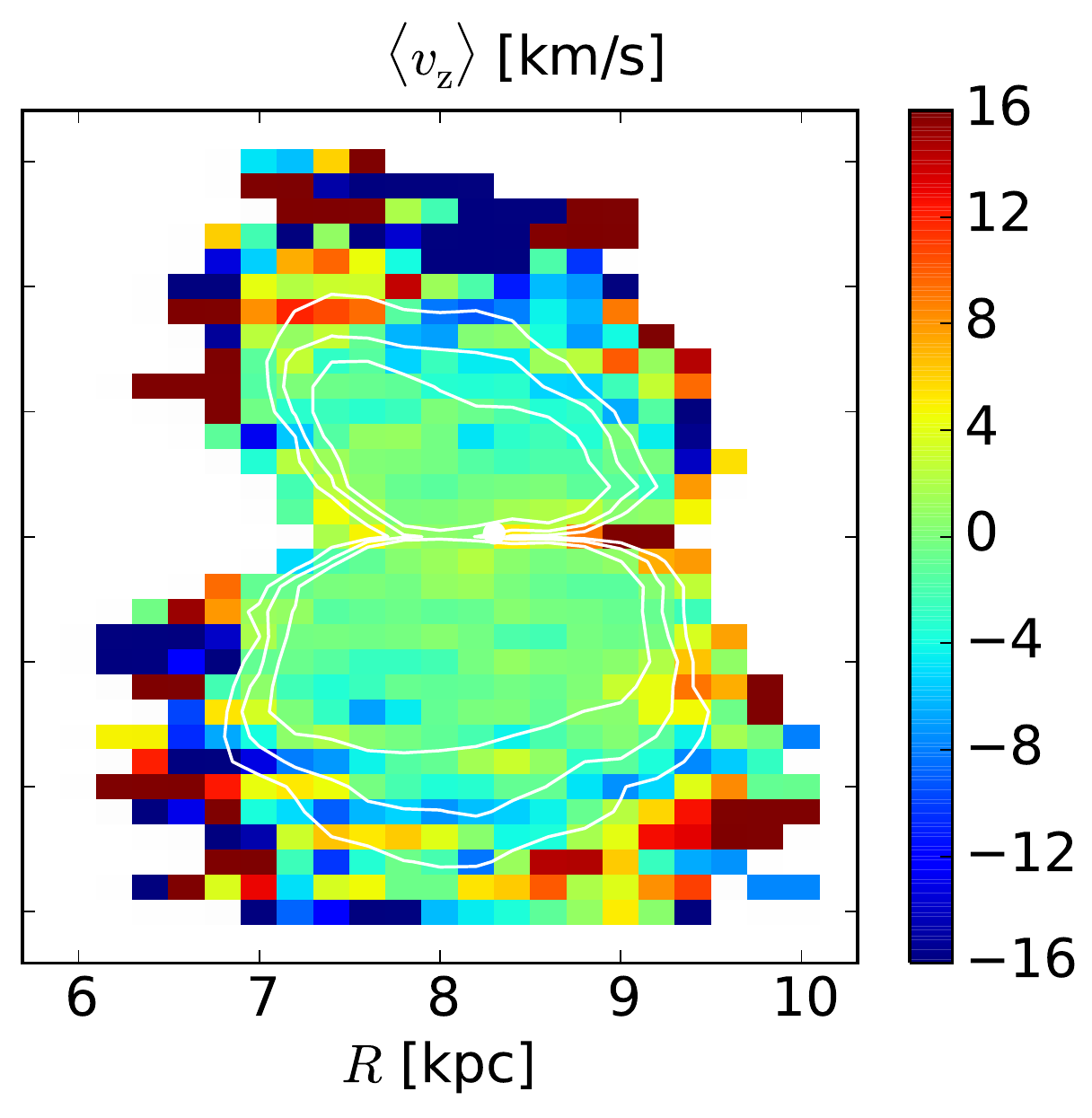}%
  \caption{The mean velocities of RC stars in the meridional plane based on the default realization in which the absolute magnitudes of all RC stars are set to $M^{\textrm{RC}}_{K_S} = -1.60$ mag. The white contours and white dot have the same meaning as in Fig. \ref{fig:Nstars_2d}.} 
  \label{fig:meanvel} 
\end{figure*}%

\begin{figure*}
  \centering
  \includegraphics[width=0.37\textwidth]{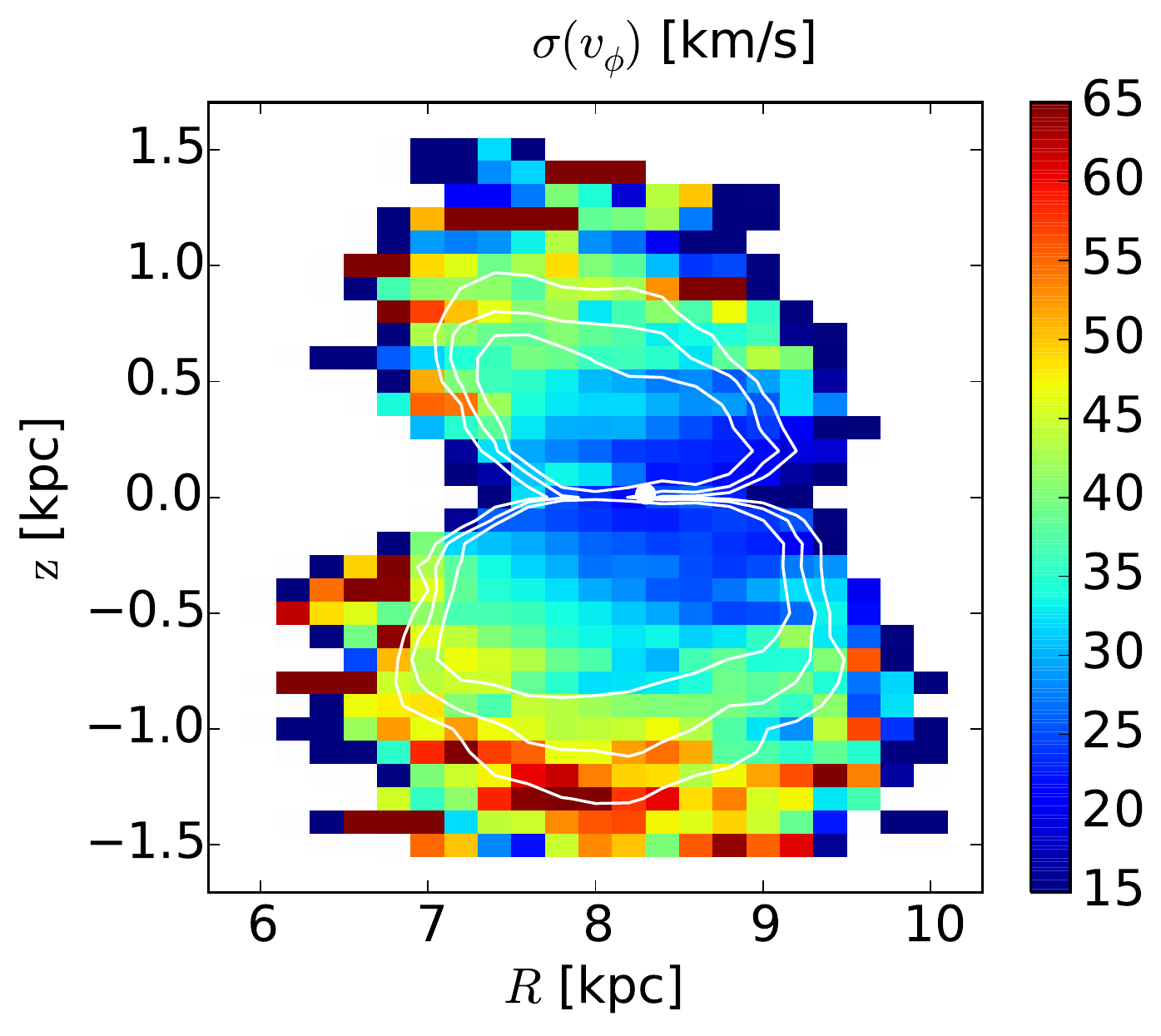}%
  \includegraphics[width=0.31\textwidth]{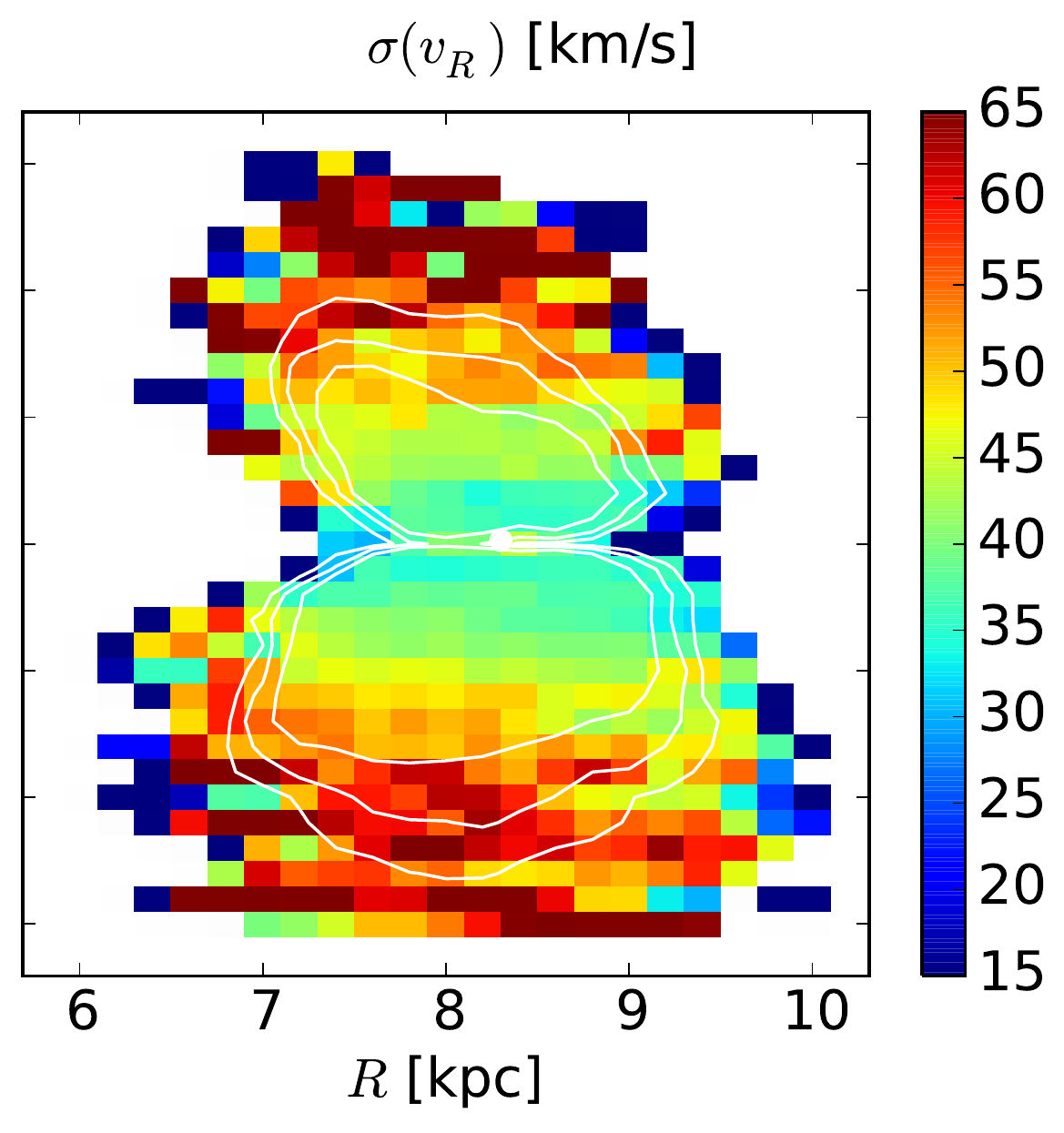}%
  \includegraphics[width=0.31\textwidth]{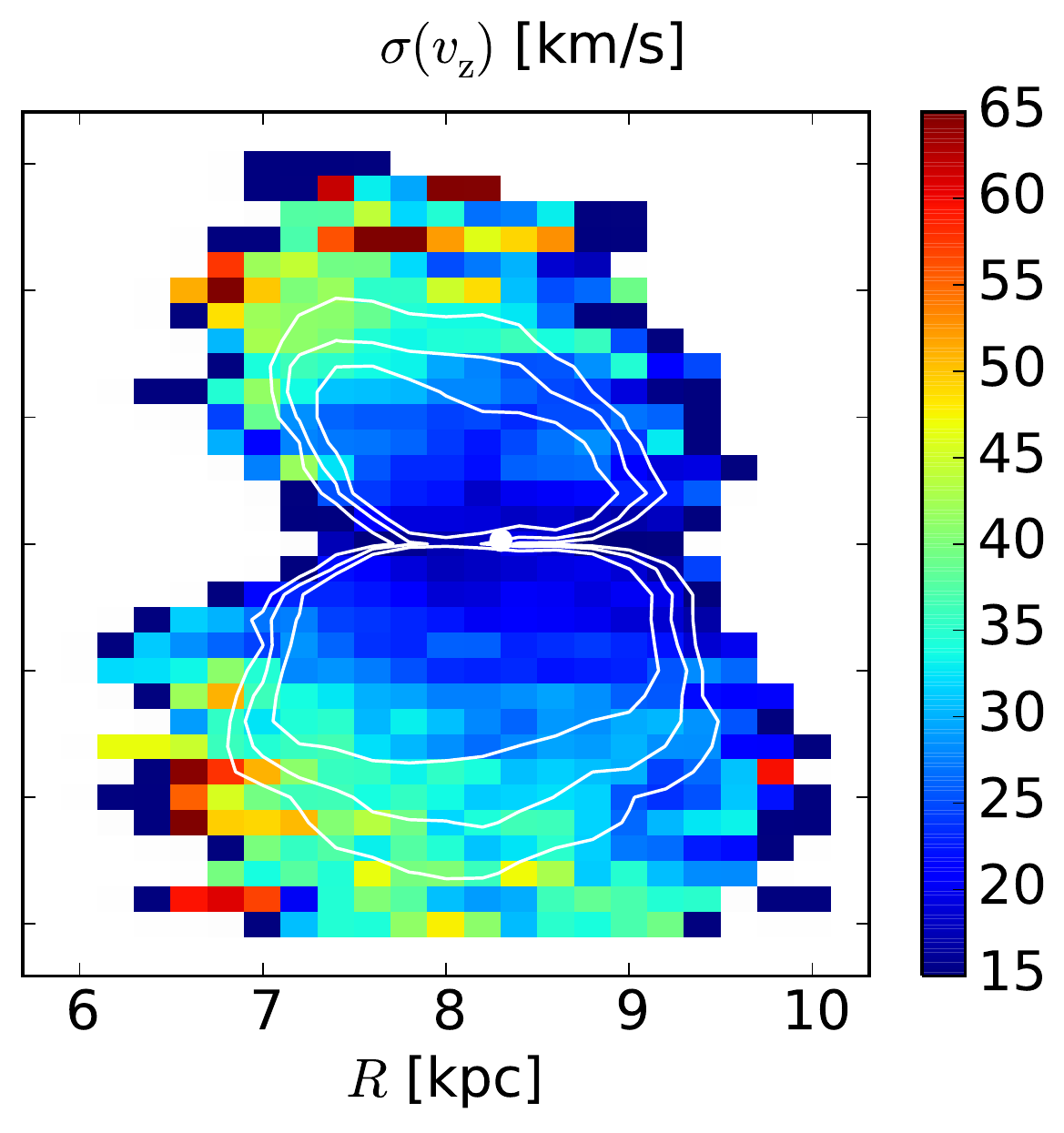}%
  \caption{Similar to Fig. \ref{fig:meanvel}, but now showing the standard deviation of the velocities.} 
  \label{fig:sigmavel} 
\end{figure*}%

When converting the observables to galactocentric cylindrical coordinates and velocities, we additionally use
\begin{itemize}
   \item $(U,V,W)_\odot = (11.1, 12.24, 7.25)$~km/s \citep{Schonrichetal2010}, for the peculiar motion of the Sun with respect to the Local Standard of Rest (LSR), and where $U$ is radially inward, $V$ in the direction of galactic rotation, and $W$ perpendicular to the Galactic plane and positive towards the north Galactic pole;
   \item $v_\mathrm{c}(R_\odot) = 240$~\text{km/s} \citep[][]{Piffletal2014}, for the circular velocity at $R=R_\odot$.
\end{itemize}
Figs. \ref{fig:meanvel} and \ref{fig:sigmavel} show kinematic maps,
i.e. mean velocities and their standard deviations, respectively, in
the meriodional plane using the same `default realization' and bin
widths as before. These figures show the well-known decrease of the
average rotational velocity $\langle v_\phi \rangle$ with height above
the plane, and the increase of the velocity dispersions with
increasing $|z|$. We also see a variation (of order 5~km/s/kpc) 
in $\langle v_R \rangle$ with respect to $R$ and some asymmetry in the vertical
direction, as found in other works \citep[e.g.][]{Casettietal2011,
  Siebertetal2011, Williamsetal2013}. On the other
hand we find no strong evidence for a bending or breathing mode in
$\langle v_z \rangle $, at least up to $0.7$ kpc \citep[see also][]{Carrilloetal2017}.




\section{The local dark matter density estimate}
\label{sec:dmdensity}

Now that we have constructed a good quality kinematic dataset, we
proceed to estimate the local dark matter density in the steady-state
axisymmetric limit. Although our kinematic maps reveal some deviations, these
are of sufficiently small amplitude\footnote{In
  the analysis carried out below we find $\langle v_R \rangle = 5.8
  \pm 3.3$~km/s and $\langle v_z \rangle = -0.6 \pm 1.9$~km/s, for the
  thick disk, and $1.4 \pm 1.1$ and $0.5 \pm 0.6$~km/s, respectively
  for the thin disk RC stars in our sample (at large heights).} that we neglect them in our analysis. In this section, we first discuss the basic
equations that relate the mass density to the kinematic moments, then
describe how we measure these moments, and finally present our new
determination and discuss the influence of the uncertainties on the
main parameters of our mass model.

\subsection{The surface mass density and the vertical Jeans equation}
\label{subsec:surfacemassdensity}

The (integrated) Poisson equation in cylindrical coordinates links the
total surface mass density $\Sigma(R,z)$ to the components of the
gravitational force per unit mass in the radial, $F_R$, and vertical
direction, $K_z$, via:
\begin{equation}
\label{eq:poisson}
-2 \pi G \Sigma(R,z) = K_z(R,z) + \int^{+z}_{0} \frac{1}{R} \frac{\partial \left(R F_R\right)}{\partial R} \mathrm{d}z^{\prime},
\end{equation}
where $\Sigma(R,z) = \int^{+z}_{-z} \rho_{\mathrm{tot}}(R,z^{\prime}) \mathrm{d} z^{\prime}$. 

Under the assumption of equilibrium, we can use the Jeans equations to relate the moments of the distribution function of a population, such as its density and velocity moments, to the gravitational potential in which it moves $\Phi(R,z)$. In this case  
\begin{equation}
\label{eq:JEQ1vertical}
\begin{split}
K_z &\equiv -\frac{\partial \Phi}{\partial z} \\
    &= - \frac{\langle v^2_{z} \rangle}{z} \left[\gamma_{\ast,z} + \gamma_{\langle v^2_{z} \rangle,z }\right] + \frac{\langle v_{R}v_{z}\rangle}{R} \left[1 - \gamma_{\ast,R} - \gamma_{\langle v_{R}v_{z} \rangle,R }\right],
\end{split}
\end{equation}
and
\begin{equation}
\label{eq:JEQ1radial}
\begin{split}
F_R &\equiv -\frac{\partial \Phi}{\partial R} \\
    &= - \frac{\langle v^2_{\phi} \rangle}{R} + \frac{\langle v^2_{R} \rangle}{R} \left[1-\gamma_{\ast,R} - \gamma_{\langle v^2_{R} \rangle,R }\right] - \frac{\langle v_{R}v_{z}\rangle}{z} \left[\gamma_{\ast,z} + \gamma_{\langle v_{R}v_{z} \rangle,z}\right] \, .
\end{split}
\end{equation}
In these equations we have defined 
$$\gamma_{Q,x} \equiv - \frac{\partial\ln \left[Q(x)\right]}{\partial\ln \left[x\right]}$$
where $\gamma_{\ast,x}$ is the log-slope of the stellar density
profile of the population with respect to coordinate $x$ (=$R$ or
$z$), and otherwise $\gamma_{Q,x}$ denotes the log-slope of the
velocity moments.  The steady state assumption implies that $\langle
v_{R}\rangle=\langle v_{z}\rangle=0$, and hence $\textrm{cov}(v_R,
v_z) = \langle v_R v_z\rangle $, and $\sigma^2(v_R) = \langle
v^2_{R}\rangle$, and analogously for $\sigma^2(v_z)$.

From Eq.~\ref{eq:JEQ1vertical} and \ref{eq:JEQ1radial} and using the
kinematic moments, we can derive the force field, which when inserted
in Eq.~\ref{eq:poisson}, allows us to derive the total surface mass
density. This will include the contributions of all baryonic
components as well as a putative dark matter component, whose
contribution we can establish with the data.  Note therefore, that not
only accurate measurements of the velocity moments and their variation
with $R$ and $z$ are needed, but also knowledge on the radial and
vertical slopes of the density distribution, as well as the surface
densities of the different baryonic components (ISM and stars). Since
our dataset does not allow us to derive these quantities reliably, we
will have to make additional assumptions.

To reduce the complexity of the problem, one may note that the last
term of Eq.~\ref{eq:poisson} is approximately zero near $R=R_\odot$
and $z=0$, since the circular velocity curve
$v_{\mathrm{c}}^2(R)=-RF_R(R,z=0)$ is approximately flat at the
solar galactocentric radius
\citep[e.g.][]{Reidetal2014}. \citet{KuijkenGilmore1989} have shown
that in this case, the term can be neglected up to a few kpc in
$z$. Furthermore, \citet{BovyTremaine2012} showed that $\frac{\partial
  \left(R F_R\right)}{\partial R}$ decreases as one moves away from
the midplane for reasonable models of the Milky Way \citep[although
this has been challenged by][]{MoniBidinetal2015}. Therefore dropping
the term will lead to an underestimate of the surface mass density
$\Sigma(R,z)$, which \citet{BovyTremaine2012} find in their models to
be a few percent at $z=1.5$ kpc up to roughly 15\% at $z=4$ kpc.

Our dataset does not extend to heights larger than $1.5$ kpc, which
implies that we can solve the integrated Poisson equation by only
evaluating the vertical Jeans Equation. We assume an exponential disk
and that $\sigma^2(v_R)$ and $\sigma^2(v_z)$ follow an exponential
profile in $R$ with the same scale length as the density
\citep[e.g.][]{vanderKruit&Searle1982,Lewis&Freeman1989}. For the tilt angle (which relates to the last term in
Eq.~\ref{eq:JEQ1vertical}), we do not assume spherical alignment\footnote{We find that the
  difference in the amplitude of $K_z$ for a spherically aligned
  ellipsoid or one where cov($v_R,v_z$) varies as $\sigma^2(v_R)$ and
  $\sigma^2(v_z)$, is less than 10\% of the uncertainty on $K_z$ due
  to measurement errors.} but that it is constant with $R$.  We then combine Eq. \ref{eq:poisson} and
\ref{eq:JEQ1vertical} to yield
\begin{equation}
\label{eq:JEQfinal}
-2 \pi G \Sigma(R,z) \simeq - \frac{\sigma(v_{z})^2}{h_z} + \frac{\partial \sigma(v_{z})^2}{\partial z} + \textrm{cov}(v_R, v_z) \left[ \frac{1}{R} - \frac{2}{h_R} \right],
\end{equation}
where $h_z$ and $h_R$ are the vertical and radial scale heights of the
population traced by the stars explored. Eq. \ref{eq:JEQfinal} can be
applied to multiple populations that satisfy the assumptions described.

We emphasize that as significantly more data with much better quality
will be available soon (e.g. Gaia DR2), one should aim to solve the
full set of equations (i.e. including all terms in Eq.~\ref{eq:poisson}), especially when going towards larger galactic
heights. Getting a better handle on both the vertical and radial slopes of the velocity moments and on the density profiles of the
samples traced would reduce the number of assumptions needed to estimate the
dark matter density.




\subsection{Analysis of the data}
\label{subsec:dataanalysis}

To mimic the uncertainties of the distances, we draw $1000$
realizations of the absolute $K_S$-band magnitude for each star in our
RC sample, assuming a Gaussian distribution characterised by our
calibrated mean and dispersion (see Sect. \ref{sec:RCsample}).  This
effectively leads to $1000$ distance realizations for the RC stars
selected from the PJM2017 sample. In each realization we transform the
observables to a galactocentric cylindrical coordinate frame and
propagate the errors. For this procedure, we assume no error on the
positional coordinates, and use the TGAS values for the proper motion
errors and their correlations.  Since radial velocities are observed
independently by RAVE there are no correlations with the proper motion
measurements from TGAS.

Given that we are only interested in the vertical trend of the
gravitational force $K_z$, in each realization we consider only those
stars that are within $0.5$ kpc in $R$ from $R_\odot$. We fold the
data below the plane towards positive $z$ by flipping the signs of $z$ and $v_z$
and select the subset of stars that trace the population
under consideration (see Sect. \ref{subsubsec:atwocomponentanalysis}). 
We then take bins in $z$ such that they contain at least 100 stars from the population
and from each bin we remove the outlier stars, by iteratively
eliminating stars outside the tilted velocity ellipsoid that would contain
99.994\% of the stars in case of a perfect multivariate Gaussian
distribution. This clipping is performed in $(v_R,v_z,v_{\phi})$-space.

Since measurement errors in general will inflate the observed velocity
dispersion, we subsequently attempt to solve for the intrinsic
velocity dispersions of the population by maximizing the
bivariate Gaussian likelihood function $L$ of the $(v_R,v_z$)-data. In
here, the error terms are set by the true intrinsic dispersions
$\sigma_\textrm{intr}$ of the population and the measurement errors
$\epsilon_i$ of star $i$, summed in quadrature:
\begin{equation}
\begin{split}
L_i &= L_i[\langle v_R \rangle, \sigma(v_R)_\textrm{intr}, \langle v_z \rangle, \sigma(v_z)_\textrm{intr}, \textrm{cov}(v_R, v_z)_\textrm{intr}] \\
& = \frac{1}{\sqrt{\textrm{det}(2 \pi \vec{\Sigma_i})}} \exp \left[
-\frac{1}{2} (\vec{x_i} -\vec{\mu})^\intercal \vec{\Sigma_i}^{-1} (\vec{x_i}-\vec{\mu})
\right] \, ,
\end{split}    
\label{eq:likelihood}
\end{equation}
where $\vec{x_i} = [v_{R,i}, v_{z,i}]$, $\vec{\mu} = [\langle v_R \rangle, \langle v_z \rangle]$, and 
\[\vec{\Sigma_i} = 
\begin{bmatrix}
    {\sigma^2(v_R)_\textrm{intr}} + \epsilon^2(v_{R,i})               & \textrm{cov}(v_R, v_z)_\textrm{intr} + \textrm{cov}(v_{R,i}, v_{z,i}) \\
    \textrm{cov}(v_R, v_z)_\textrm{intr} + \textrm{cov}(v_{R,i}, v_{z,i}) & {\sigma^2(v_z)_\textrm{intr}} + \epsilon^2(v_{z,i})
  \end{bmatrix}, \]
  with $\textrm{cov}(v_{R,i}, v_{z,i})$ the covariance of the
  measurement errors in the $(v_{R}, v_{z})$ components for star
  $i$. Thus 
\begin{equation}
\label{eq:likelihood2}
    L = \prod^{N}_{i=1} L_i \, ,
\end{equation}
given the $N$ stars of the population in the bin under consideration. 

We employ MCMC modelling \citep{Foreman-MacKeyetal2013} to solve for the intrinsic velocity dispersions $\sigma(v_R)_\textrm{intr}$ and $\sigma(v_z)_\textrm{intr}$, the mean velocities $\langle v_R \rangle$ and $\langle v_z \rangle$, and the covariance term $\textrm{cov}(v_R, v_z)_\textrm{intr}$ in each bin in $z$. Each MCMC run also returns an error estimate of these moments. Priors to the MCMC model are added such that the dispersions in $v_R$ and $v_z$ should be positive and such that the absolute value of the correlation of $v_R$ and $v_z$ is always smaller or equal to $1$. 

The procedure of taking bins in $z$, removing outlier stars, and solving for the intrinsic moments is repeated for every realization of the RC sample and for each population of stars.

\begin{figure}[t] 
\centering
 \includegraphics[width=8.5cm]{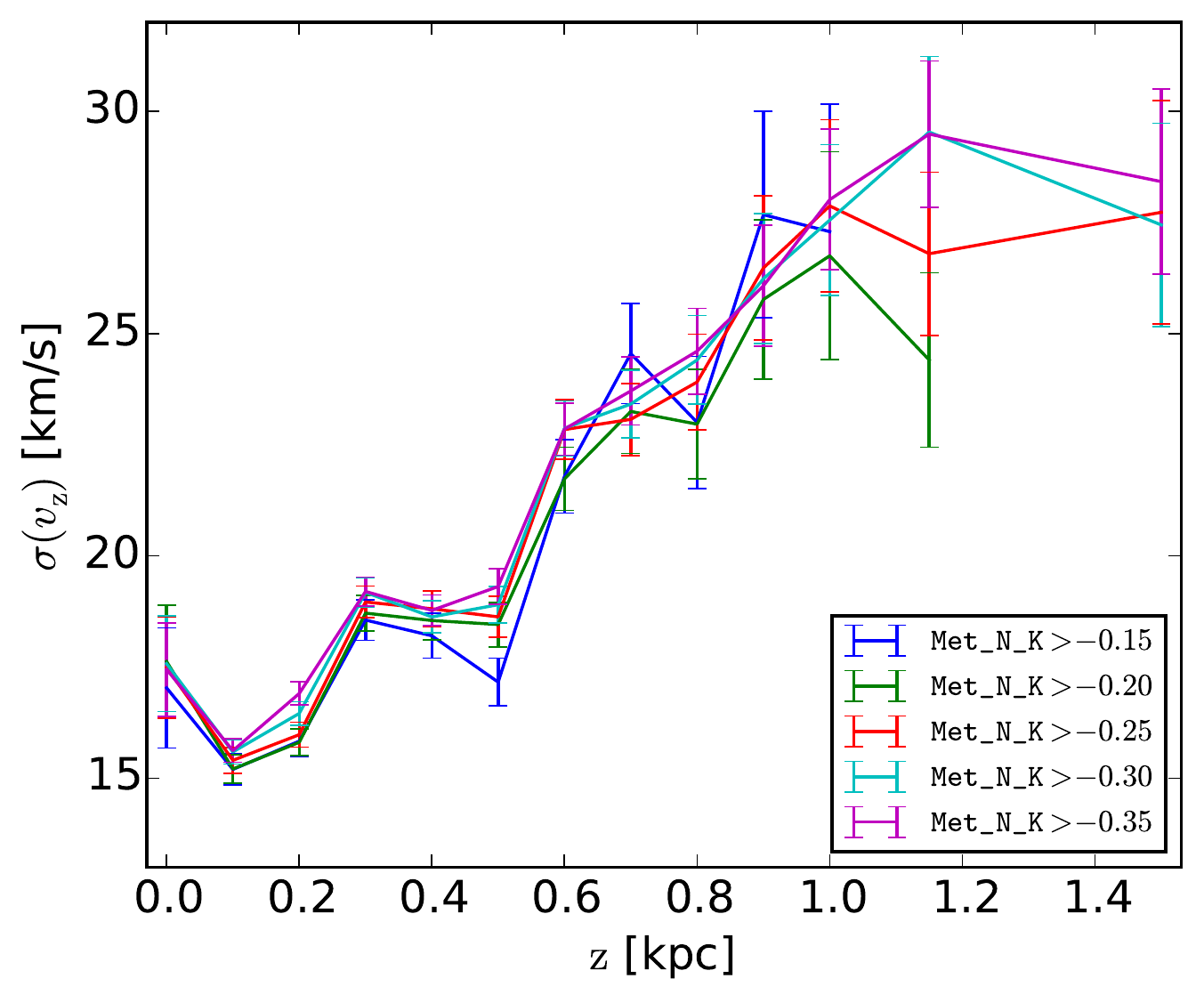}\\
 \includegraphics[width=8.5cm]{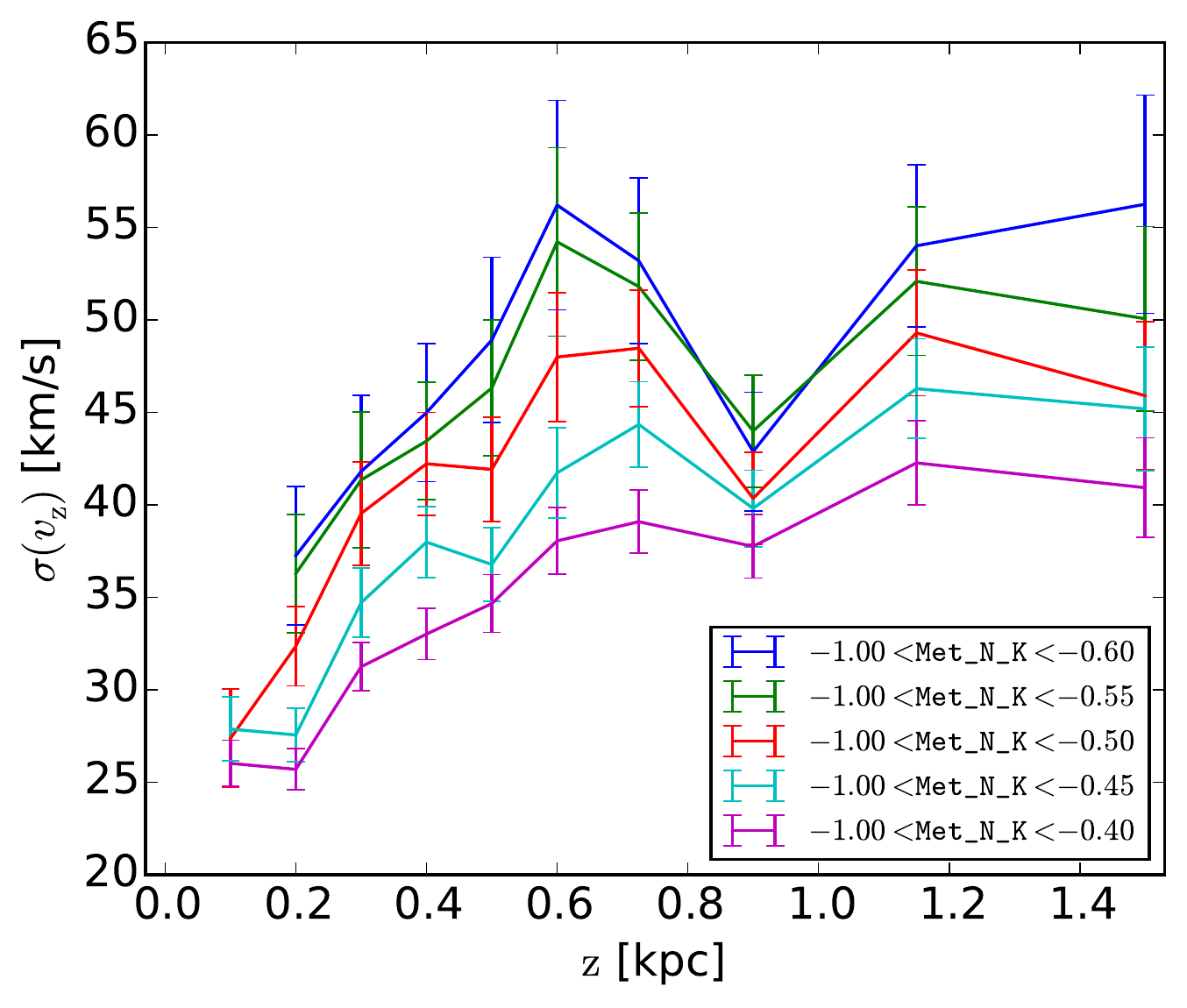}%
 \caption{Influence of the adopted metallicity ranges for the thin
   (top) and thick (bottom) disk RC samples on the intrinsic vertical
   velocity dispersions (for the `default' realization). Only for the
   thick disk velocity dispersion profile there is a dependence on 
   the adopted upper metallicity boundary, partly due to contamination
   by the low metallicity tail of the thin disk. The bins plotted here
   are fully independent (non-overlapping).}
 \label{fig:RCmetallicityboundaries} 
\end{figure}%




\begin{figure} \centering 
  \includegraphics[width=8.5cm]{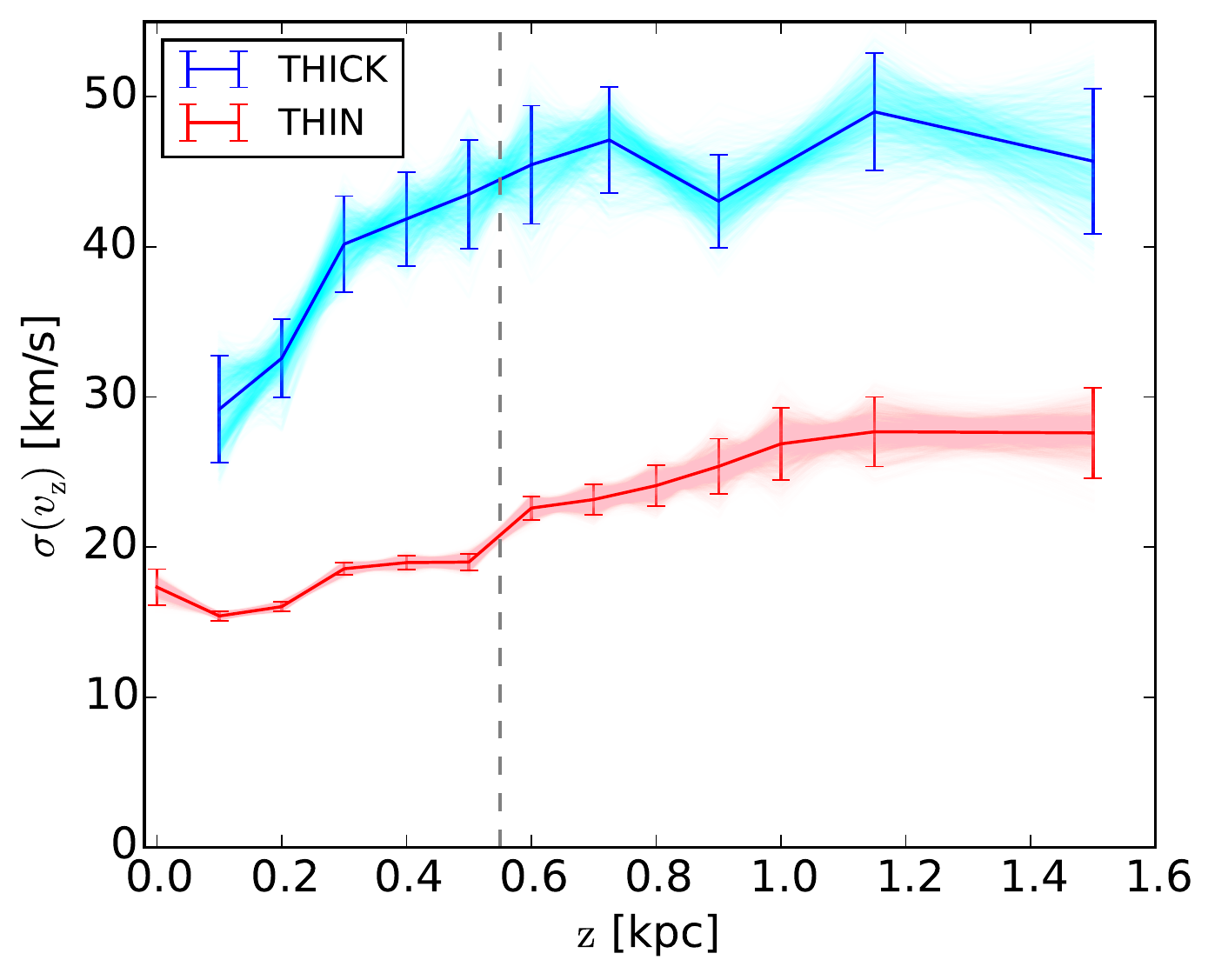} 
  \caption{Mean intrinsic vertical velocity dispersion profile over
    all realizations for both the thin metal-rich (red) and thick
    metal-poor (blue) disk RC stars. The error bars show the spread
    over all realizations due the spread in the absolute magnitude of
    the RC stars and as a result of the error deconvolution.}
 \label{fig:dataanalysis_sigmaz} 
\end{figure}

\subsubsection{Vertical velocity dispersion profiles: Two components}
\label{subsubsec:atwocomponentanalysis}

The Milky Way contains a thin and a thick disk that, in the Solar
neighbourhood, are known to have different spatial, kinematical and metallicity
distributions. In particular, the thick disk is hotter, has a larger
scale height and is typically more metal-poor, becoming more dominant
for [Fe/H]$\lesssim -0.5$~dex. These two populations should thus be
treated separately when attempting to solve Eq.~\ref{eq:JEQfinal}.

Since the distributions of thin and thick disk stars show a certain
amount of overlap in most observables, we first investigate how the
choice of different metallicity ranges for the thin and thick disk
tracer RC samples affect the intrinsic $z$-velocity dispersion
profiles. For the metallicity we use the calibrated values,
$\texttt{Met\_N\_K}$, as provided by RAVE.  In
Fig. \ref{fig:RCmetallicityboundaries} we show the results using the
`default realization' (i.e. for the distance) after deconvolution of
the measurement errors (see Eqs. \ref{eq:likelihood} and
\ref{eq:likelihood2}). For this figure we require a minimum of 50
stars per bin.

This figure shows that a change in the adopted metallicity boundary
for the thin disk has no significant effect on the vertical velocity
dispersion profile (top panel). Therefore, for the thin disk we simply
choose $\texttt{Met\_N\_K} \geq -0.25$ dex.  For the thick disk on the
other hand, changing the upper metallicity limit can have a
significant influence on the recovered dispersion profile (see the bottom
panel of the figure), most likely driven by contamination by thin disk
stars. To get a clean sample (and to avoid the inclusion of halo
stars) we adopt the following range in what follows: $-1.00 \leq
\texttt{Met\_N\_K} \leq -0.50$ dex. Although we could have chosen an
even lower value for the upper limit, this comes at the cost of an
important decrease in the number of stars. For example in the default 
realization there are 586, 427, or 386 stars in the thick disk sample 
in between $z=0.55$ and $z=1.7$~kpc 
when respectively using upper metallicity boundaries of $-0.5$, $-0.55$, and $-0.6$~dex.

After finding the intrinsic velocity moments as function of $z$ for
each of the $1000$ realizations for the samples representing the thin and
thick disk, we compute their mean and dispersion over all
realizations, also taking into account the errors on the moments in
each single realization (which were determined in the MCMC modelling
procedure). These dispersions thus account for the error on the
moments due to the errors in proper motion and radial velocity and to
the unknown distances to the RC stars. In
Fig. \ref{fig:dataanalysis_sigmaz} we show the mean vertical velocity
dispersion profiles for both the thick (blue) and thin (red) disk
sample, overplotted on the dispersion profiles from all individual
realizations (for which the error bars are omitted in this plot). From
this figure we see that the velocity dispersion profiles increase
quickly with $z$ and that above $z \sim 0.5$~kpc their variation is
much shallower, particularly for the thick disk sample.

\subsubsection{The $K_z$ force}
\label{subsubsec:Kzforce}
Because our goal is to measure more reliably the contribution of dark
matter we focus on the bins at large galactic heights, since for small
$z$ the baryons (are expected to) dominate the gravitational force. We
thus choose to explore only those bins for which the central
$z$-coordinate satisfies $|z|\geq 0.6$ kpc. 

We are now ready to insert the moments (and their errors) in Eq. \ref{eq:JEQfinal}. 
This equation also requires knowledge of the
variation of $\sigma(v_z)$ with $z$. Rather than differentiating the
data directly, we fit a linear function to $\sigma(v_z)$ as function
of $z$, whose slope and uncertainty are then used in Eq. \ref{eq:JEQfinal}.  

To compute $K_z$, we set $R=R_\odot$ in the last term of
Eq. \ref{eq:JEQfinal}, and we fix the scale lengths of both disks to
an intermediate value of $2.5$ kpc \citep[e.g.][]{Siegeletal2002,
  Juricetal2008, Bensbyetal2011, McMillan2011, Bovyetal2012,
  Robinetal2012, Robinetal2014, Bland-Hawthorn&Gerhard2016}. Finally
we explore how $K_z$ varies for a set of scale heights for the thin
and thick disk populations.

\begin{figure} 
\centering 
  \includegraphics[width=8.5cm]{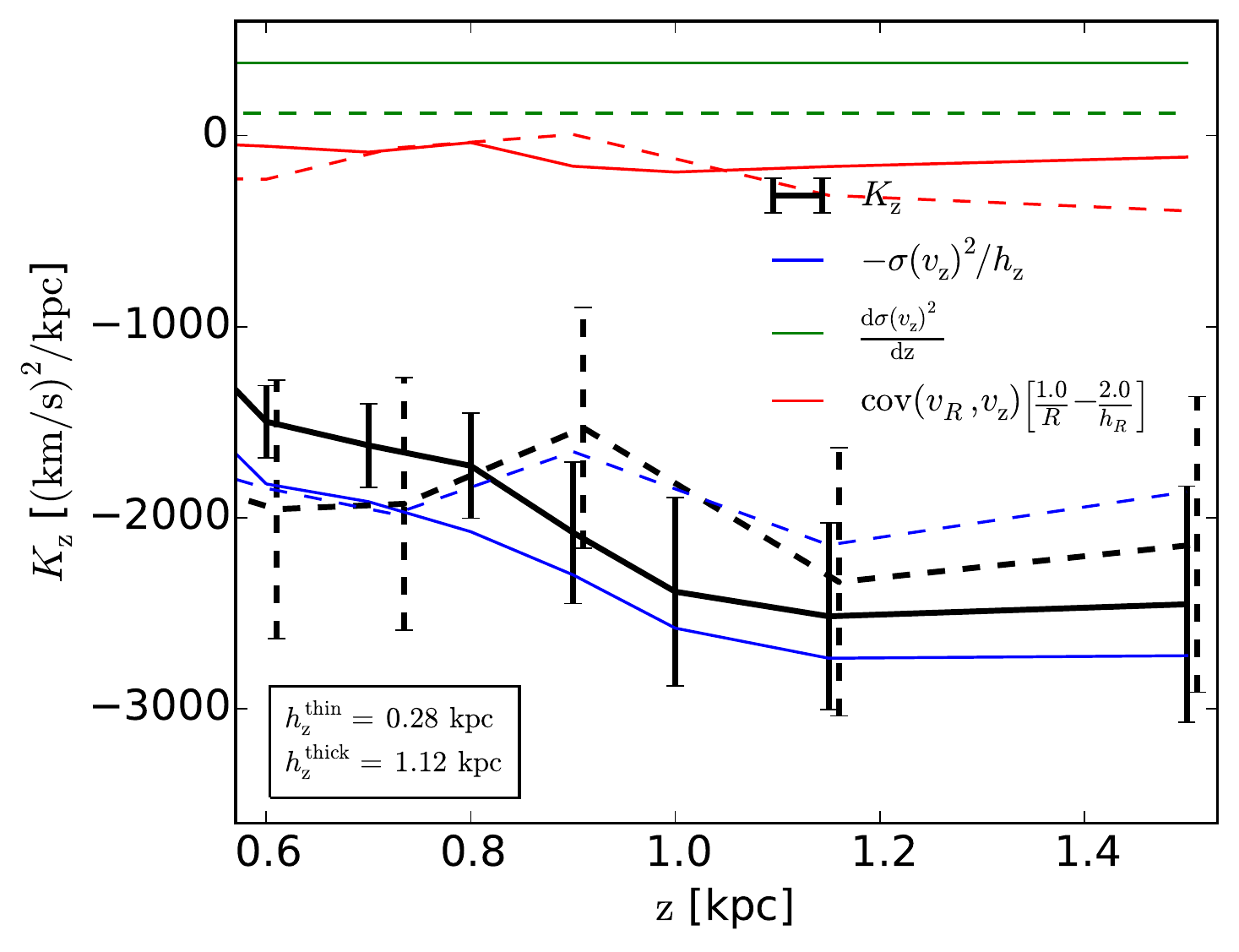}%
  \caption{Derived vertical force $K_z$ (black lines) for the
    thin (solid lines) and the thick disk samples (dashed lines). The
    coloured lines show the decomposition of $K_z$ into the terms
    concerning the vertical velocity dispersion (blue), the slope of
    the vertical variance as function of $z$ (green), and the mixed
    velocity moment (red). The curves shown here were computed for the
    combination $h^{\mathrm{thin}}_{z} = \hZthinbest$ kpc and
    $h^{\mathrm{thick}}_{z} = \hZthickbest$ kpc.}
 \label{fig:dataanalysis_kz} 
\end{figure}

Fig. \ref{fig:dataanalysis_kz} shows the derived vertical forces in
black for $h^{\mathrm{thin}}_{z} = \hZthinbest$~kpc and
$h^{\mathrm{thick}}_{z} = \hZthickbest$~kpc. The solid line
corresponds to the thin disk sample, the dashed line to the thick disk
sample.  Since we folded our data towards positive $z$, the forces
derived are negative.  We find $K_z^{\rm thin} = -2454 \pm 619$ and $K_z^{\rm
  thick} = 2141 \pm 774$~$(\mathrm{km/s})^2 \! /\mathrm{kpc}$ 
  at 1.5 kpc away from the Galactic plane. We also show the decomposition of $K_z$ into the three
terms of Eq. \ref{eq:JEQfinal}: the first associated to vertical
velocity dispersion (blue), the second to the slope of the vertical
variance as function of $z$ (green), and the third to the mixed
velocity moment (red). The first term dominates $K_z$ for both
samples. On the other hand we find that the error on $K_z$ is
dominated by that on the vertical velocity dispersion for the thin
disk sample, while for the thick disk it is the error on the slope.

\begin{figure*}[t] \centering
  \includegraphics[width=0.485\textwidth]{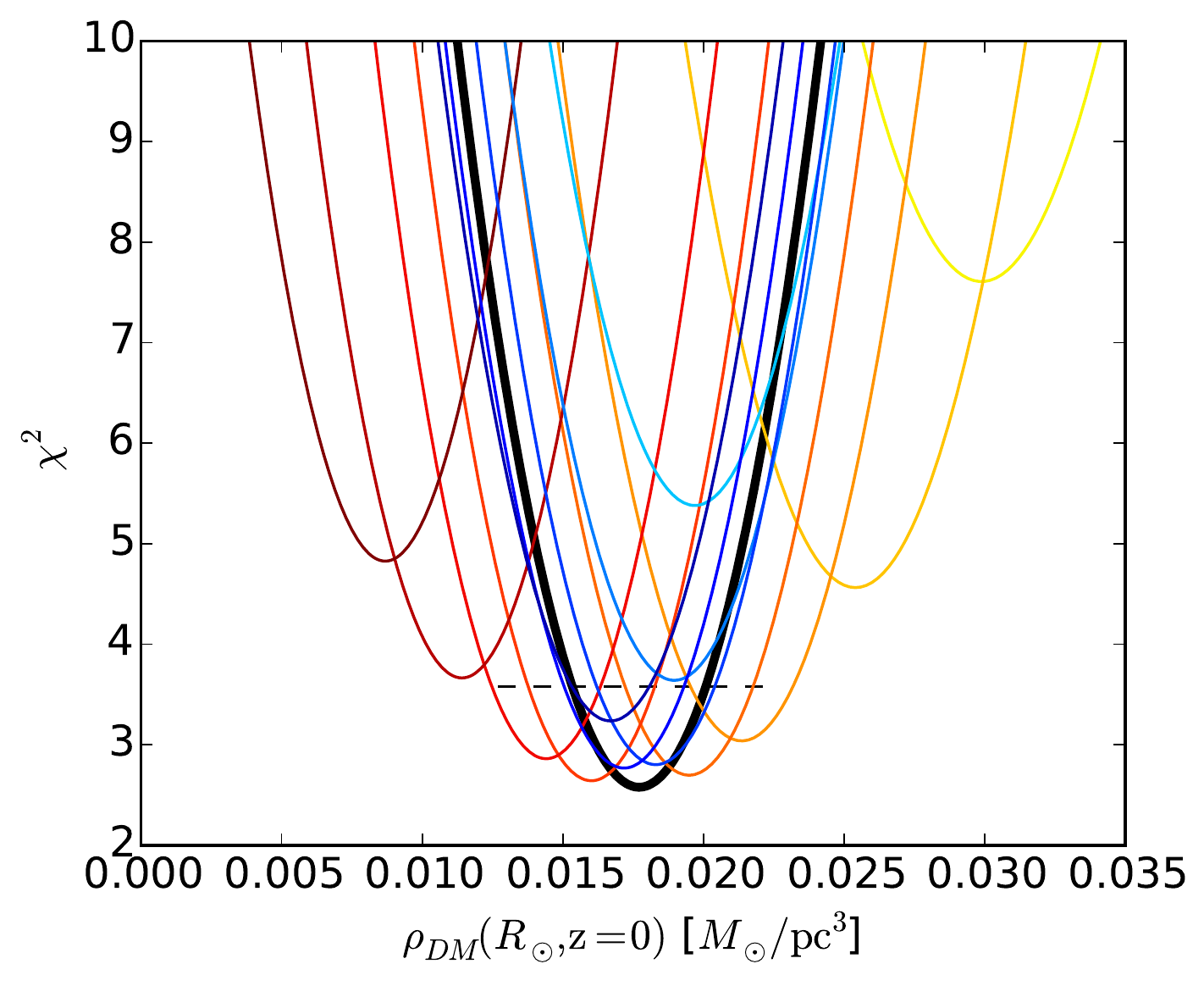}%
  \includegraphics[width=0.505\textwidth]{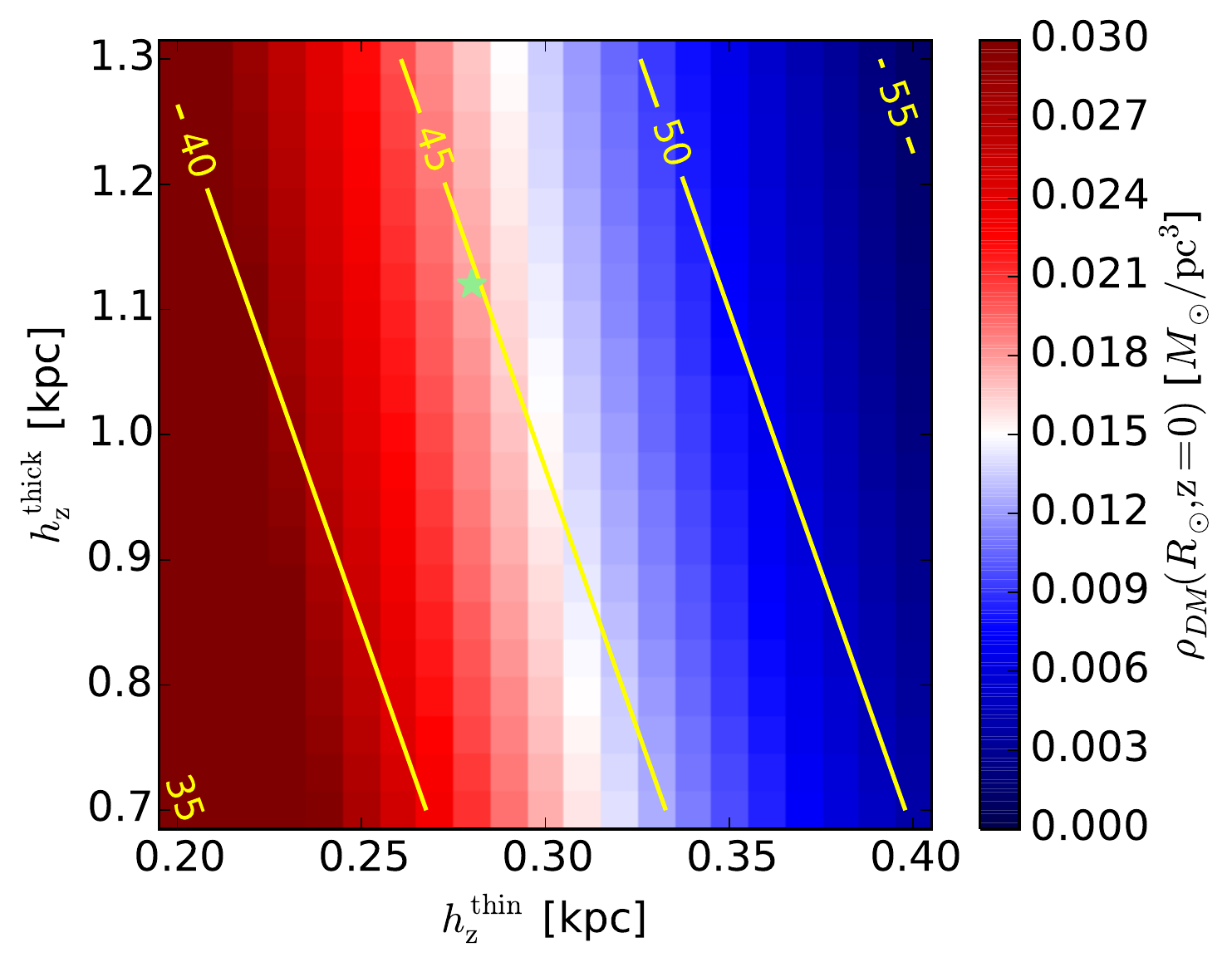}%
  \caption{Left: $\chi^2$ values as function of the local dark matter
    density, for different combinations of the scale heights of the
    thin and thick disks. The (light to dark) blue curves have
    $h_z^\textrm{thin} = \hZthinbest$ kpc and
    $h_z^\textrm{thick}=\left[0.85,0.94,1.03,1.21,1.30\right]$~kpc,
    while those from orange to dark red have
    $h_z^\textrm{thin}=\left[0.22,0.24,0.26,0.27,0.29,0.30,0.32,0.34\right]$~kpc
    and $h_z^\textrm{thick} = \hZthickbest$ kpc. The thick black curve
    corresponds to the global minimum of the $\chi^2$, and which is
    found for model with $h_z^\textrm{thin} = \hZthinbest$~kpc and
    $h_z^\textrm{thick} = \hZthickbest$~kpc. The horizontal dashed
    line indicates a $\Delta \chi^2 = 1$ with respect to this minimum
    $\chi^2$-value. Right: Map showing the dependence of the best-fit
    values of the local dark matter density on the adopted scale
    height for the thin (x-axis) and thick (y-axis) disk. The total
    $\chi^2$ is minimum at the position of the star symbol. The yellow
    lines are drawn where the total baryonic surface mass density in
    the models equals $35,40,45,50$ and $55 M_\odot$/pc$^2$.}
  \label{fig:rhoDM} 
\end{figure*}%

\subsection{The mass model}
\label{subsec:massmodel}

Now that we have determined $K_z$, and hence the total surface mass
density $\Sigma(z)$ according to Eq.~\ref{eq:poisson} (which is
strictly equivalent only if we assume $\partial (RF_R)/\partial R=0$ for all
$z$), we proceed to compare it to the surface mass density derived from
the contribution of all baryonic components and explore the need for
dark matter.

We characterize the baryonic components with a double exponential
stellar thin and thick disk, and an infinitely thin ISM disk with a
surface mass density equal to $13 M_\odot$/pc$^2$
\citep{Holmberg&Flynn2000}, and consider a constant dark matter
density\footnote{The assumption of a constant dark matter density is a
  reasonable approximation for the volume probed by our data, since
  for example the difference in surface mass density with an NFW halo with a
  scale radius of 14.4~kpc and a
  flattening $0.9$ as in \citet{Piffletal2014}, is only $\sim1\%$ at $z=1.5$~kpc.}. We further fix the midplane stellar
density at the solar radius to $0.043 M_\odot$/pc$^3$
\citep{McKeeetal2015}, which is very similar to the value of $(0.040
\pm 0.002) M_\odot$/pc$^3$ found by \citet{Bovy2017}. We fix the
thick to thin disk density ratio to $12\%$ \citep{Piffletal2014}.

To explore the effect of the assumed thin and thick disk scale heights
on the local dark matter density estimates, we sample the thin disk
scale height from 200 to 400 pc and the thick disk scale height from
700 to 1300 pc, such that typical estimates reported in the literature are covered
\citep[e.g.][]{Bland-Hawthorn&Gerhard2016}. For each combination we determine the value of $\rho_{DM}$
by minimizing the total $\chi^2$-value defined as:
\begin{equation}
    \label{eq:chi2}
    \chi^2 = \sum^{N_{\rm thin}}_{j=1} \left( \frac{\textrm{model}(j) - \textrm{data}(j)}{\textrm{error}(j)} \right)_{\rm thin}^2 +  \sum^{N_{\rm thick}}_{j=1} \left( \frac{\textrm{model}(j) - \textrm{data}(j)}{\textrm{error}(j)} \right)_{\rm thick}^2\,
\end{equation}
where $j$ runs over all datapoints for which surface mass densities
have been derived for both the thin and thick disk data. The best fit
model will thus minimize jointly $\chi^2_{\rm thin} +
\chi^2_{\rm thick}$, but to understand the quality of the models, we
inspect separately the reduced $\chi^2_{\rm thin,red}$ and $\chi^2_{\rm
  thick,red}$.


The results for different combinations of scale height parameters are
shown in Fig. \ref{fig:rhoDM}. In the left panel we show the
$\chi^2$-values as function of the local dark matter density parameter
for a number of models. The thick black curve shows the case of a
model with scale heights $h_z^\textrm{thin}=\hZthinbest$ and
$h_z^\textrm{thick}=\hZthickbest$ kpc. This is the model with the
lowest $\chi^2$ (among all different combinations explored).  The
dashed horizontal line corresponds to $\Delta \chi^2 = 1.0$ with
respect to its minimum value, the criterion we use to quantify the
error on the dark matter density determination. In this panel we also show (from orange to dark red) the $\chi^2$
curves for models with varying thin disk scale height for fixed
$h_z^\textrm{thick}=\hZthickbest$~kpc. Models with fixed
$h_z^\textrm{thin}=\hZthinbest$ kpc, but with varying thick disk
scale height are given with colours varying from light to dark
blue. We thus see that a relatively large change in thick disk scale
height does not alter the inferred dark matter density by much,
although $\Delta \chi^2 > 1$ for e.g. $h_z^\textrm{thick} \le
0.95$~kpc. On the other hand, a small change in the thin disk scale
height has a large influence on the inferred dark matter density.

This is also depicted in the right-hand side panel of
Fig. \ref{fig:rhoDM}, where we show the best-fit dark matter densities
for all combinations of scale heights explored. The green star marks
the location in $(h_z^\textrm{thin}, h_z^\textrm{thick})$-space where
the model can fit the data the best. The larger impact of the thin
disk scale height on the fitted dark matter density is probably driven
by the smaller error bars on the gravitational force profile implied
from the thin disk sample in comparison to that from the thick disk
sample (see e.g. Fig.~\ref{fig:dataanalysis_sigmaz}). Yellow lines
indicate where the total baryonic surface mass density in the models
equals $35,40,45,50$ and $55 M_\odot$/pc$^2$. Most previous works seem
to agree on baryonic surface mass densities in the range $40-55
M_\odot$/pc$^2$ \citep[e.g.][and references therein]{McKeeetal2015},
thus the models with combinations of scale heights which result in
smaller or larger baryonic contributions are less plausible.

\begin{figure}[t] 
\centering
 \includegraphics[width=4.3cm]{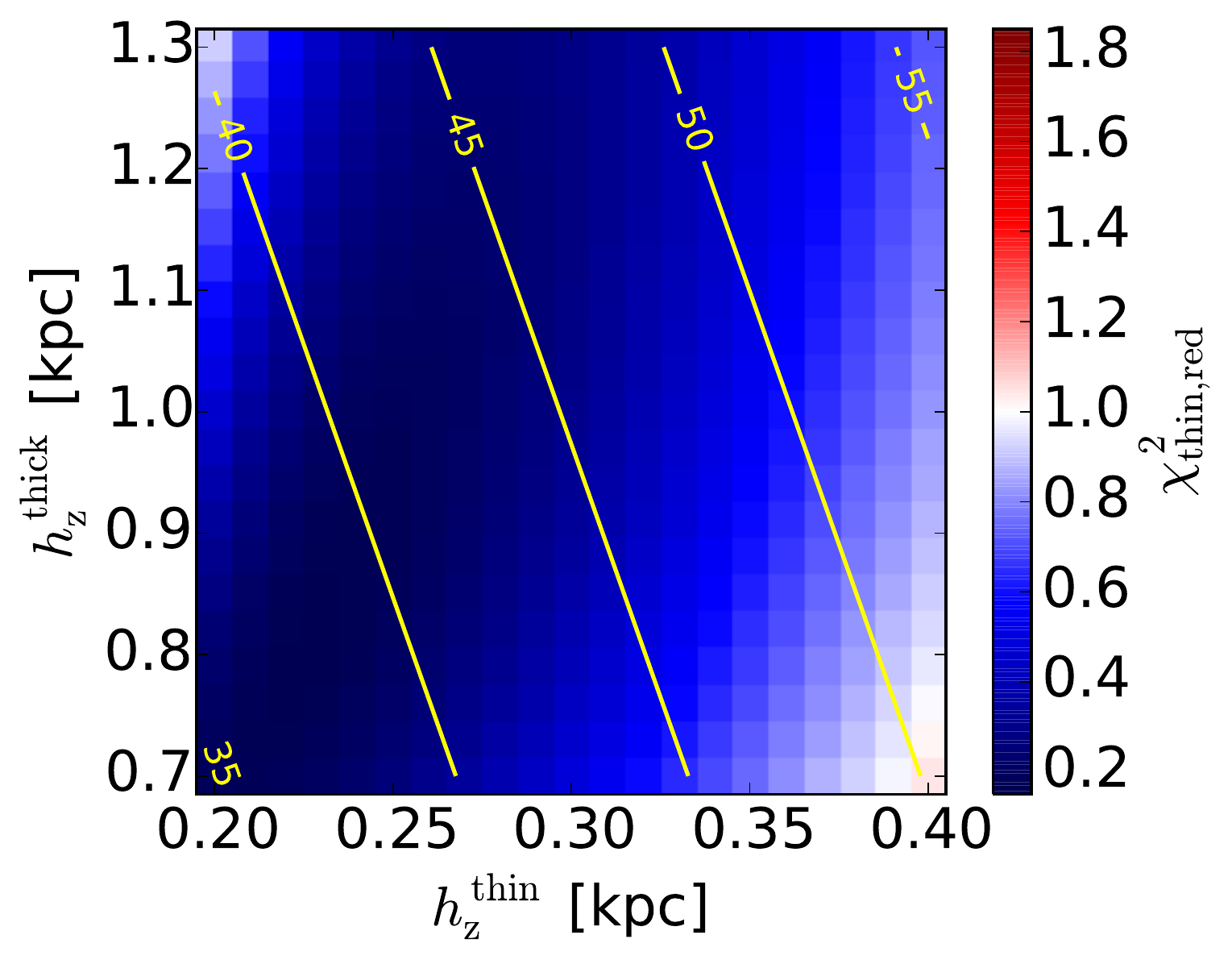}
 \includegraphics[width=4.3cm]{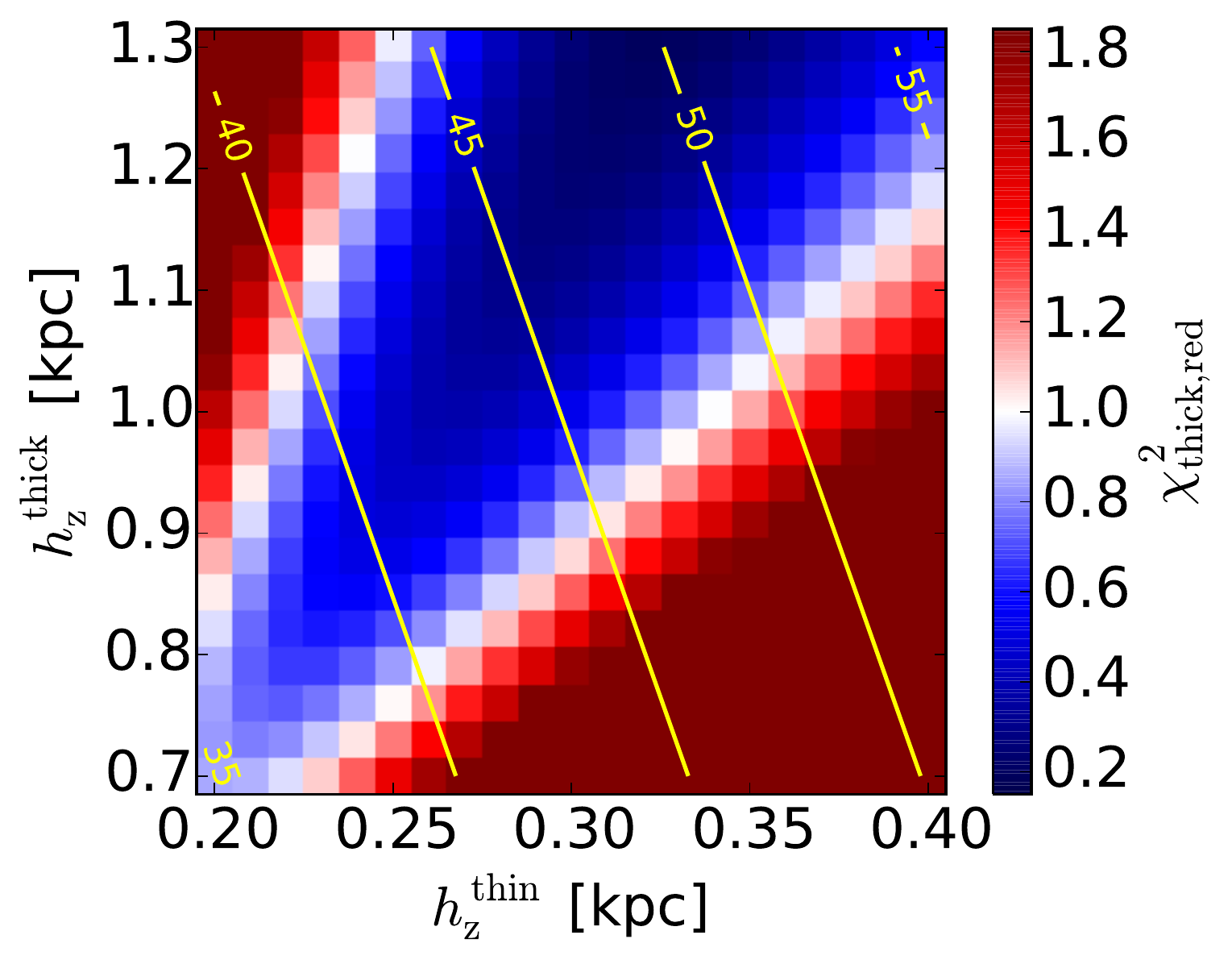}%
 \caption{Maps of the reduced $\chi^2$-values computed for the thin
   (left) and thick (right) disk samples separately. The thin disk
   sample is generally fitted well with the minimization of the joint
   $\chi^2$, and this is because it contributes with a larger number
   of datapoints with smaller error bars. This is not the case for the
   thick disk, where for certain combinations of the scale
   heights the fit is poor for this sample ($\chi^2_{\rm thick, red} > 1$).}
 \label{fig:chi2thinthick} 
\end{figure}

In Fig. \ref{fig:chi2thinthick} we explore further the quality of the
fits obtained by comparing the reduced $\chi^2$-values as computed
from the thin (left) and the thick disk data
(right). The panels show that the thin disk data is always fitted
well, but that models with a very low (or large) thin and large (or
low) thick disk scale height do not lead to a high quality fit for the
thick disk data.

\begin{figure} \centering 
 \includegraphics[width=8.5cm]{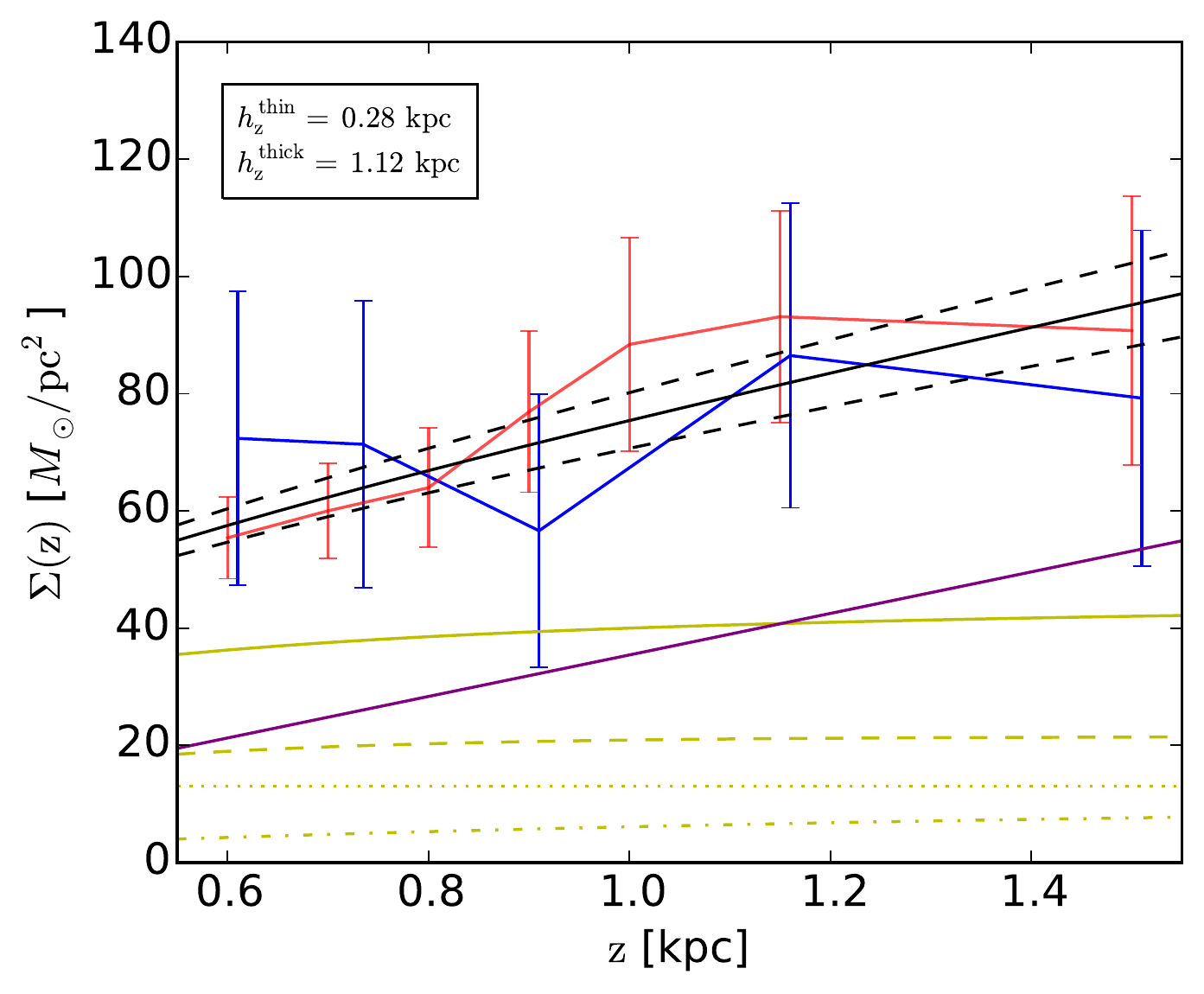}
 \caption{The surface mass density as implied from the axisymmetric
   Jeans Equation, derived for the thin (red) and thick (blue) disk
   samples. In this example the scale height for the thin disk tracer
   stars (red) is set to $\hZthinbest$ kpc, the scale height for the
   thick disk tracer stars (blue) to $\hZthickbest$ kpc. This
   combination of parameters gives the lowest $\chi^2$-value (see
   Fig. \ref{fig:rhoDM}) when fitting our mass model (black), which is
   the sum of the baryonic (solid yellow) components (thin, ISM and
   thick disks from top to bottom in dashed yellow) and the dark
   matter component (purple). }
 \label{fig:massmodelexample} 
\end{figure}

In summary, the model and data are thus most consistent for
$h_z^\textrm{thin} = \hZthinbest$ kpc and $h_z^\textrm{thick} =
\hZthickbest$ kpc, which results in $\Sigma_{\mathrm{baryon}}
=\Sigmabaryonbest M_{\odot}/\textrm{pc}^2$. In this case, the inferred
local dark matter density is $\rhodmbest \pm \erhodmbest
M_{\odot}/\textrm{pc}^3$. This model is shown in
Fig. \ref{fig:massmodelexample} and has $\chi^2 = \chisqbest$. With
$11$ degrees of freedom ($12$ datapoints and $1$ model parameter),
this value implies that the quality of the fit is good. Note that the
uncertainty on the scale heights of the populations used constitute a
much larger source of error on the estimate of the local dark matter
density than the measurement errors alone, which lead to a relative
error of order 10\% only.

In the analysis presented thus far we have considered all data within
$|R - R_\odot| \le 0.5$~kpc. When we restrict ourselves to a smaller
volume, within $0.25$~kpc from $R_\odot$, the dataset is smaller and
this leads to less constraining power, especially for the thick disk
set, on the value of the local dark matter density. For this more
local sample, we find that the model does not fit the thin disk data
well ($\chi_{\rm thin, red}^2 > 1$), for $h_z^\textrm{thin} >
0.35$~kpc, and for the thick disk if $h_z^\textrm{thick} \lesssim 3.3
h_z^\textrm{thin}$. 


\section{Conclusions}
\label{sec:conclusions}

We have studied the kinematics of red clump stars from the RAVExTGAS
dataset. The kinematic maps obtained are well behaved, do not show
evidence for strong bending or breathing modes up to $0.7$ kpc from
the midplane, and only show a small hint of radial motions towards the
outer disk. We therefore applied the axisymmetric Jeans Equation
relating kinematic moments to the vertical force, ultimately yielding
a new measurement of the dark matter density in the Solar
Neighbourhood. To account for the presence of multiple populations, we
divided the red clump sample into thin and thick disk tracer samples
according to the metallicity of the stars, as estimated from the RAVE
dataset, and fitted both populations simultaneously. This allowed us
to determine a local value of the dark matter density with a relative
internal error (due to measurement errors on the observables) of only
13.5\%, \mbox{$\rho_\textrm{DM}(R_\odot,0) = \rhodmbest \pm
  \erhodmbest$ $M_\odot/\textrm{pc}^3$}, in reasonable agreement with
previous work.

It is however misleading to consider only the internal errors on the
dark matter density, as they do not account for the large systematic
uncertainties in the stellar disk parameters (especially the scale
heights), the thick-to-thin disk density ratio, the scale lengths of
the disks, and the ISM mass. These (external) sources of uncertainty
can lead to a large systematic error on the dark matter density near
the Sun. For example a 10\% difference in the scale height of the thin
disk leads to a 30\% change in the value of $\rho_\textrm{DM}$, and a
nearly equally good fit to the data. The change due to uncertainties
on the scale height of the thick disk is slightly weaker. It is
therefore extremely important to get accurate constraints on the
stellar disk parameters of the tracer stars used. 


\begin{acknowledgements}
We thank M.A. Breddels, D. Massari, L. Posti and J. Veljanoski for numerous discussions. A.H. acknowledges financial support
from a VICI grant from the Netherlands Organisation for Scientific Research,
N.W.O. 
This work has made use of data from the European Space Agency (ESA)
mission {\it Gaia} (\url{https://www.cosmos.esa.int/gaia}), processed by
the {\it Gaia} Data Processing and Analysis Consortium (DPAC,
\url{https://www.cosmos.esa.int/web/gaia/dpac/consortium}). Funding
for the DPAC has been provided by national institutions, in particular
the institutions participating in the {\it Gaia} Multilateral Agreement.
Funding for RAVE has been provided by: the Australian Astronomical Observatory; the Leibniz-Institut fuer Astrophysik Potsdam (AIP); the Australian National University; the Australian Research Council; the French National Research Agency; the German Research Foundation (SPP 1177 and SFB 881); the European Research Council (ERC-StG 240271 Galactica); the Istituto Nazionale di Astrofisica at Padova; The Johns Hopkins University; the National Science Foundation of the USA (AST-0908326); the W. M. Keck foundation; the Macquarie University; the Netherlands Research School for Astronomy; the Natural Sciences and Engineering Research Council of Canada; the Slovenian Research Agency; the Swiss National Science Foundation; the Science \& Technology Facilities Council of the UK; Opticon; Strasbourg Observatory; and the Universities of Groningen, Heidelberg and Sydney.
The RAVE web site is at https://www.rave-survey.org.

\end{acknowledgements}



\bibliographystyle{aa}
\bibliography{bibliography}

\end{document}